\documentclass{emulateapj}
\usepackage[utf8]{inputenc}
\usepackage{graphicx}
\usepackage{color}
\usepackage[dvipsnames]{xcolor}
\usepackage{amsmath}

\newcommand{\msun}{\ensuremath{\,M_\odot}}
\newcommand{\ma}{\ensuremath{M_{\rm a}}}
\newcommand{\mb}{\ensuremath{M_{\rm b}}}
\newcommand{\mbh}{\ensuremath{M_{\rm BH}}}
\newcommand{\mzams}{\ensuremath{M_{\rm ZAMS}}}
\newcommand{\mzamsa}{\ensuremath{M_{\rm ZAMS,a}}}
\newcommand{\mzamsb}{\ensuremath{M_{\rm ZAMS,b}}}

\newcommand{\azams}{\ensuremath{a_{\rm ZAMS}}}

\newcommand{\rsun}{\ensuremath{\,R_\odot}}
\newcommand{\zsun}{\ensuremath{\,Z_\odot}}

\newcommand{\yr}{\ensuremath{\,\mathrm{yr}}}
\newcommand{\kyr}{\ensuremath{\,\mathrm{kyr}}}
\newcommand{\myr}{\ensuremath{\,\mathrm{Myr}}}
\newcommand{\gyr}{\ensuremath{\,\mathrm{Gyr}}}

\newcommand{\kms}{\ensuremath{\,\mathrm{km}\,\mathrm{s}^{-1}}}
\newcommand{\ergs}{\ensuremath{\,\mathrm{erg}\,\mathrm{s}^{-1}}}

\newcommand{\startrack}{{\tt StarTrack}}
\newcommand{\dash}{\,\hbox{--}\,}
\newcommand{\sci}[2]{\ensuremath{#1\times10^{#2}}}

\newcommand{\rSBHa}{\ensuremath{\mathcal{R}_{\rm dSBH,1}}}
\newcommand{\rSBHb}{\ensuremath{\mathcal{R}_{\rm dSBH,2}}}
\newcommand{\rlmxba}{\ensuremath{\mathcal{R}_{\rm MTBHB,1}}}
\newcommand{\rlmxbb}{\ensuremath{\mathcal{R}_{\rm MTBHB,2}}}
\newcommand{\rdcoa}{\ensuremath{\mathcal{R}_{\rm DCO,1}}}
\newcommand{\rdcob}{\ensuremath{\mathcal{R}_{\rm DCO,2}}}
\newcommand{\rmer}{\ensuremath{\mathcal{R}_{\rm mSBH}}}

\newcommand{\std}{STD}
\newcommand{\midz}{mid-$Z$}
\newcommand{\lowz}{low-$Z$}
\newcommand{\ssa}{SS0}

\newcommand{\nkr}{NK$_{\rm R}$}
\newcommand{\nkbe}{NK$_{\rm BE}$}
\newcommand{\imff}{flat IMF}%$\Gamma=-1.9$}
\newcommand{\imfs}{steep IMF}%$\Gamma=-2.7$}

\newcommand{\ivbh}{\ensuremath{v_\mathrm{BH,0}}}

\begin{document}

\title[Populations of stellar mass Black holes]{Populations of stellar mass Black holes from binary systems}

\author{Grzegorz Wiktorowicz\altaffilmark{1,2,3}\thanks{E-mail: gwiktoro@astrouw.edu.pl},
        \L{}ukasz Wyrzykowski\altaffilmark{3},
        Martyna Chruslinska\altaffilmark{4},
        Jakub Klencki\altaffilmark{4},
        Krzysztof A. Rybicki\altaffilmark{3},
        Krzysztof Belczynski\altaffilmark{5}}

 \affil{  
     $^{1}$ National Astronomical Observatories, Chinese Academy of Sciences, Beijing 100101, China\\
     $^{2}$ School of Astronomy \& Space Science, University of the Chinese Academy of Sciences, Beijing 100012, China\\
     $^{3}$ Astronomical Observatory, Warsaw University, Al. Ujazdowskie 4, 00-478 Warsaw, Poland\\
     $^{4}$ Institute of Mathematics, Astrophysics and Particle Physics, Radboud University Nijmegen, PO Box 9010, 6500 GL\\
     $^{5}$ Nicolaus Copernicus Astronomical Center, Polish Academy of Sciences, Bartycka 18, 00-716 Warsaw, Poland \\
            }

\begin{abstract}
	In large and complicated stellar systems like galaxies it is difficult to predict the number and characteristics of a black hole (BH) population. Such populations may be modelled as an aggregation of homogeneous (i.e. having uniform star formation history (SFH) and the same initial chemical composition) stellar populations. Using realistic evolutionary models we predict the abundances and properties of BHs formed from binaries in these environments. We show that the BH population will be dominated by single BHs (SBH) originating from binary disruptions and stellar mergers. Furthermore, we discuss how BH populations are influenced by such factors as initial parameters, metallicity, initial mass function (IMF), and natal kick (NK) models. As an example application of our results, we estimate that about $26$ microlensing events to happen every year in the direction of the Galactic Bulge due to BHs in a survey like OGLE-IV. Our results may be used to perform in-depth studies related to realistic BH populations, e.g. observational predictions for space survey missions like Gaia, or Einstein Probe.	We prepared a publicly available database with the raw data from our simulations to be used for more in-depth studies.
\end{abstract}


\section{Introduction}

A BH is defined as a region in space from which nothing, even light, can escape \citep[for a recent review see][]{Bambi1711}. BHs may be detected when interacting with other objects \citep[e.g. in X-ray binaries; ][]{Remillard0609,Corral-Santana1603}, as gravitational wave sources \citep[e.g.,][]{Abbott1602,Abbott1706}, when they interfere with radiation \citep[e.g. as gravitational lenses; ][]{Wyrzykowski1605}, or in non-interacting binaries by observations of their companions \citep[e.g.][]{Thompson1806,Mashian1709}. It was also proposed that SBHs may be observed as X-ray novae \citep[e.g.][]{Matsumoto1803}.

The so called stellar-mass BHs, i.e. stellar-origin BHs and BHs originating from mergers of stars and/or stellar-origin compact objects, with masses from $\sim5$ to possibly {\it a few }$\times100\msun$ are final stages of massive stars evolution \citep[e.g.][]{Neugebauer03} and are distinguished from other subgroups such as super-massive BHs \citep[$\mbh\gtrsim10^6\msun$; e.g.][]{Ferrarese0502}, intermediate-mass BHs \citep[$\sim10^2<\mbh<\sim10^5\msun$; e.g.][]{Mezcua2017, Koliopanos1801} and hypothetical primordial BHs \citep[e.g.][]{Chapline1975, Khlopov1006}. Hereafter, we focus exclusively on stellar-mass BHs.

\citet{Corral-Santana1603} provided a list of 59 BHs in transient X-ray binaries (XRB). 22 of them have dynamically confirmed mass estimates \citep[see][for a recent list]{Casares1701}. 5 BHs were detected in High-mass XRBs (defined as heaving a donor mass above $10\msun$). Recently, gravitational waves made it possible to discover first double BH merger events \citep{Abbott1602,Abbott1606,Abbott1706,Abbott1710}, and a double NS merger GW170817, which may have formed a low-mass BH \citep{Pooley1712}. Interestingly, \citet{Adams1707} observed a $\sim25\msun$ star to disappear after a short brightening, which may be interpreted as a BH-formation event. Up to now, none SBHs were confirmed \citep[e.g.][]{Tsuna1801}.

Previous studies of BH populations are usually outdated and do not take into account recent progress in our understanding of massive star evolution \citep[e.g.,][]{Langer1209,Vink1503,Vink15}. The earliest estimations predicted $\sim100$ million BHs in the Milky Way (MW) galaxy \citep{Shapiro83} with as much as $45\%$ of supernovae occurring in close binaries \citep{Tutukov9202}. \citet{Timmes9602} provided an upper limit of $1.4\times10^9$ and \citet{Samland9803} estimated $1.8\times10^8$ BHs in the galaxy taking into account  changes in star formation rate (SFR). \citet{Belczynski0203,Belczynski0408} used the \startrack\ population synthesis code in its earlier version to investigate BHs formed in star formation (SF) bursts. Their results indicate that most of BHs in the MW are actually single objects (not in binaries), whereas these which remain bound will have mostly main-sequence (MS) companions, provided that the stellar population is not very old. 

Recent studies do not provide the estimations for parameter distributions of the entire BH population. For example, \citet{Elbert1801} predicted $\sim10^8$ BHs in a MW-type galaxy, however, their approach does not take into account binary interactions directly and focuses on binary BHs (BH+BH) only. \citet{Lamberts1801} performed a detailed cosmological simulation of the MW evolution (including e.g. changes in metallicity) and predicted, that $\sim1$ million of BH binaries have already merged in the galaxy, whereas $\sim3$ millions are still present. However, their study does not account for BHs with lower-mass companions, or disrupted systems, which may actually constitute a bulk of the BH population \citep{Belczynski0408}.

Although the current population of BHs is predicted to be mainly single, their progenitors did not have to be single stars. Due to a NK \citep[e.g.][]{Lyne9405,Janka1309}, a binary (especially with large orbital separation, or a low-mass companion) may be disrupted \citep{Iben9712}. Recent observational studies \citep[see e.g. ][for a review]{Sana1711} suggest that even $\gtrsim90\%$ of massive stars (BH progenitors) are born in binaries (or multiple systems). The presence of a companion may significantly affect the evolution of a progenitor and, therefore, final properties of a BH. Some studies used a simplified approach. E.g., \citet{Elbert1801} encapsulated all the interactions in just two parameters. In this study we focused on a detailed consideration of binary interactions, however, we neglected any higher-order systems \citep[e.g.][]{Toonen1612}.

A serious problem in obtaining estimations of BH populations is the lack of knowledge about the stellar environment. Taking the MW as an example, not even the total stellar mass is well constrained, let alone the SFH, or the distribution and motions of stars. What is more, a galaxy with complicated evolution and structure cannot be modelled with a uniform population of objects evolved from a single SF burst. Therefore, in our approach we take into account a range of evolutionary models with different initial parameters. Such results can be joined together according to SFH and chemical evolution of a stellar population in order to obtain more realistic distributions.  

Specifically, we propose a step-by-step approach in which we use results from modelling of homogeneous and scalable BH populations in order to build a complicated (non-homogeneous) stellar systems like galaxies. We provide the estimates for a simplified MW galaxy as an exemple (for detailed modeling of the MW galaxy see Olejak et al. in prep.). However, our results may be used to simulate virtually any large stellar population providing the dynamical interactions between stars may be neglected (it is not true e.g. for globular clusters).

The two most important factors that influence BH populations are the metallicity, and NKs. The metallicity of a star significantly affects the mass loss in stellar wind and, therefore, the final compact object's mass \citep{Fryer0106}. \citet{Belczynski1005} showed that the maximal BH mass for solar metallicity environment is $\sim15\msun$, whereas for lower metallicity environments may reach $\sim80\msun$. However, their study did not consider the pair-instability supernova and pair-instability pulsation supernova \citep[e.g.][]{Woosley1702,Woosley1901} and was performed for single-stars only. Both, pair-instability supernovae and pair-instability pulsation supernova are believed to produce the second "mass gap" in the distribution of BH masses between $\sim50\dash135\msun$ \citep{Belczynski1610,Spera1710,Marchant1810,Woosley1901}. However, in binary systems the masses of BHs can fill this range due to mass transfer (MT) phases and mergers \citep[e.g.][]{Spera1905}.

Uncertainties in the modelling of common envelope (CE) phase add additional significant error to the predictions of binary evolution \citep[for a review see][]{Ivanova1302}. The CE seems to be essential for the formation of XRBs where a strong reduction of initial separation is necessary to produce a Roche lobe overflow (RLOF), or significant accretion from wind. However, it was shown \citep{Wiktorowicz1409}, that different prescriptions for the CE phase, give similar results with respect to the population of XRBs. In contrast, the formation of close double compact objects (DCO), with time-to-merger smaller than the Hubble time, which also typically involves the CE phase, is highly influenced by the adopted CE model \citep[e.g.][]{Dominik1211}. Nonetheless, as we show in this paper, only a small part of the total BH population resides in XRBs and close DCOs. Most stellar mergers of isolated binaries occur as a result of failed CE ejection, therefore, CE model may strongly influence the population of BHs originating form merger products. However, even though outcomes of mergers may constitute a significant fraction of a BH population, the poorly understood physics of stellar mergers and post-merger evolution make it impossible to reliably describe the population of these objects. In this paper, therefore, we do not include the analysis of CE models.

The fraction of binaries that remain bound after a supernova (SN) explosion as well as peculiar velocities of BHs are particularly sensitive to the assumptions about the NKs. The former impacts the predicted ratio of the number of SBHs to those found in binaries. The latter influences the spacial distribution of BHs in stellar systems (e.g. galaxies). Although the SN mechanism is not well understood, NKs are usually being connected to asymmetries arising in this process. Main candidates are: mass ejection \citep[e.g.][]{Wongwathanarat1304}, gravitational waves \citep{Bonnell9503}, and neutrinos \citep[e.g.][]{Fryer0604}. \citet{MillerJones1403} derived peculiar velocities of several BH binaries (BHB, i.e. binaries composed of a BH and a non-compact companion) obtaining values between $19\div144\kms$. However, it is usually difficult to assess NKs from present-day motions \citep{Fragos0906,Repetto1210,Repetto1511}. Recent analysis performed by \citet{Repetto1705} has shown that at least some of the BHs in the MW should have obtained high NKs ($\sim100$ km/s), comparable to these of NSs. On the other hand, \citet{Mandel1602} showed that velocities higher than $\sim80$ km/s are not necessary, although cannot be ruled out. Additionally, \citet{Jonker0410} showed that XRBs with BH accretors have similar spacial distribution as theses with NS accretors, what suggests a similar NK magnitude at birth. See \citet{Belczynski1603} and references therein for a recent discussion on the BH NKs.

In this paper, we performed the first simulation of the entire BH population for different evolutionary models. Our analysis is focused on SBHs from binary disruption events (Sec.~\ref{sec:SBH}), BHBs, DCOs (Sec.~\ref{sec:bhb}), and potential BHs originating from stellar mergers (Sec.~\ref{sec:mergers}). We pay a particular attention to the description of the results for the standard model (\std), which may depict the MW disk field population, and differences in relation to other models.

The main goals of this work are: 1) to describe the general characteristics of the BH population and its dependence on the most important model parameters (e.g., metallicity, NKs), and 2) to provide a reference point for forthcoming in-depth studies focusing on astrophysical problems involving BH populations like: XRBs, microlensing, gravitational wave sources, to name a few. Especially educating are possible comparison studies with the results of actual and future surveys focused on gravitational wave sources (e.g., aLIGO, Einstein Telescope, LISA), X-ray sources (e.g., XMM-Newton, Chandra, NuSTAR), or gravitational microlensing (OGLE, Gaia, LSST), which may result in obtaining better constrains for the evolutionary models. 

Particularly, in Sec.~\ref{sec:ml} we estimate a number of microlensing events toward the Galactic bulge. The microlensing method seems to be a promising way to search for SBHs, which, as we show in this work, constitute a vast majority of all BHs in stellar populations.

In order to allow for easier and more flexible usage of the data introduced by this paper, we prepared a publicly available database where all the results are available for download: \begin{center}\it https://universeathome.pl/universe/bhdb.php\end{center}

\section{Methodology}\label{sec:methods}

We utilize a recent version of the \startrack\ population synthesis code \citep[with further updates]{Belczynski0206,Belczynski0801}. The code has been frequently used to study BHs in XRBs and DCOs \citep[e.g.][]{Dominik1211,Dominik1312,Dominik1506,Wiktorowicz1409,Belczynski1610,Klencki1708}. Recent updates include, but are not limited to, new prescriptions for wind mass loss from massive stars \citep{Vink1111}, pair-instability supernovae and pair-instability pulsation supernovae \citep{Belczynski1606,Woosley1702}.

\begin{deluxetable}{cl}
    \tablewidth{\columnwidth}
    \tablecaption{Summary of models}

    \tablehead{ Model & difference in respect to standard model }
    \startdata
    \std & standard (reference) model: \\
     & \hspace{0.5cm}solar metallicity \zsun\\
     & \hspace{0.5cm}distribution of initial periods\\
     & \hspace{0.5cm}\hfill $P(\log P)\sim(\log P)^{-0.55}$\\
     & \hspace{0.5cm}distribution of initial eccentricities\\
     & \hspace{0.5cm}\hfill $P(e)\sim e^{-0.42}$\\
     & \hspace{0.5cm}BH/NS natal kicks are drawn from \\
     & \hspace{0.7cm}Maxwellian distribution with $\sigma=265$ km s$^{-1}$\\
     & \hspace{0.7cm}BH natal kicks reduced due to fallback\\
     & \hspace{0.5cm}moderate slope for high-mass end of the IMF\\
     & \hspace{0.5cm}\hfill $\Gamma=-2.3$\\
    \midz & metallicity equal to $10\%\zsun$\\
    \lowz & metallicity equal to $1\%\zsun$\\
    \ssa & distribution of initial separations $\log(a)\sim1$\\
     & distribution of initial eccentricities $P(e)\sim e$\\
    \nkr\ & BH NKs are inversely proportional to the BH's mass \\
    & \hfill (Eq.~\ref{eq:NK_R})\\
    \nkbe & BH/NS natal kick proportional to \\
     & \hfill ratio of ejecta mass and remnant mass (Eq.~\ref{eq:NK_BE})\\
    \imff & flat slope for high-mass end of the IMF ($\Gamma=-1.9$)\\
    \imfs & steep slope for high-mass end of the IMF ($\Gamma=-2.7$)\\
    \enddata
    \tablecomments{List of models. All the main parameters are provided only for \std\ model. For other models only differences in respect to \std\ model are given explicitly.}
\label{tab:models}
\end{deluxetable}

Here, we analyse 8 main models (Tab.~\ref{tab:models}) which differ in parameters that significantly affect the resulting BH population. Additionally, in the web database, we provide a grid of all 54 models which allow to investigate combinations of the main models. For each model we have simulated the evolution of $2\times10^6$ binaries from zero-age main-sequence (ZAMS) to disruption, merger, or reaching the age of $15\gyr$. For mergers that are not DCOs we apply a simple formalism to estimate the endpoint of post-merger single-star evolution (see Sec.~\ref{sec:methods_merger}).

Initial primary masses (\ma) are drawn from a broken power-law distribution \citep{Kroupa9306} with $\Gamma=-1.3$ for $\ma<0.5\msun$ and $\Gamma=-2.2$ for $0.5<\ma<1.0\msun$. For initial masses $\ma>1.0\msun$ we chose $\Gamma=-2.3$ \citep[\std;][]{Kroupa01}, $-2.7$ \citep[\imfs\ model;][]{Kroupa0312}, or $-1.9$ \citep[\imff\ model;][]{Schneider1801}. Conclusions of \citet{Schneider1801} were recently revised by \citet{Farr1807} who argued in favor of a much steeper $\Gamma\approx-2.11$, or $-2.15$. Being aware of this, we leave the original value in order to emphasize the influence of IMF steepness on the population of BHs. The mass ratio $q=\mb/\ma$ (where \mb\ is the mass of the secondary -- initially less massive star) is drawn from a uniform distribution between $0.08 \msun/\ma$ and $1$. In \std\ model, we assume that the initial distribution of periods is $P(\log P)\sim(\log P)^{-0.55}$, and distribution of eccentricities is $P(e)\sim e^{-0.42}$ \citep{Sana1207}. Even though, the results of \citet{Sana1207} are for stars with initial masses between $15\div60\msun$, we extrapolate them to entire investigated range ($0.08\div150\msun$). Additionally, in model \ssa, we test a distribution of initial separations that is flat in logarithm \citep{Abt83} with maximal value set as $10^5\rsun$ and a thermal distribution of eccentricities \citep[$P(e)\sim e$;][]{Duquennoy9108}.

The effects of pair-instability supernovae and pair-instability pulsation supernovae \citep{Woosley1702} may significantly alter the final BH mass \citep[e.g.][]{Belczynski1606}. The instability leads to a significant mass loss for massive stars \citep{Woosley0711}, or disruption of the entire star \citep{Heger0203}. Following \citep{Belczynski1606}, we assume that for stars that form massive helium cores ($M_{\rm He}=45\dash65\msun$) all the envelope above the inner $45\msun$ is lost due to pulsations. Furthermore, we assume that stars which form heavier helium cores ($M_{\rm He}=65\dash135\msun$) are subject to the pair-instability supernovae and leave no remnant.

The CE phase is important for the evolution of many binaries and may lead to the formation of a much closer system, or a merger. In a general situation, we utilize the simple energy balance \citep{Webbink8402}. However, the donor type influences significantly the binary survival during this phase \citep[e.g.][]{Belczynski0706}. MS and Hertzsprung gap (HG) donors lack a clear core-envelope structure \citep[no clear entropy jump; e.g.][]{Taam0001,Ivanova0402}, so the orbital energy is transferred to the entire star, instead of being transferred to the envelope only, what prevents envelop ejection and leads to a merger. It is not clear when the core-envelop boundary emerges (late HG, or red giant phase), thus two models for the treatment of HG donors in CE were introduced \citep[e.g.][]{Dominik1211}. First one (model B), assumes that a CE phases with a HG donor always results in a merger. Second one (model A), producing significantly higher DCO merger rates \citep[e.g.][]{Belczynski0706,Dominik1211} allows for survival in such situations. We adopt model B in the present study and analyse the influence of model A in Sec.~\ref{sec:modelsAB}.

We test models with 3 different metallicities. Most of the models have solar metallicity \citep[$Z=0.02=\zsun$, where \zsun\ is the solar metallicity;][]{Villante1405}. We also introduce two models with lower metallicity: \midz\ model ($Z=0.002=10\%\zsun$), and  \lowz\ model ($Z=0.0002=1\%\zsun$). Additionally, three different NK models are included in our simulations. Those models are described below.

\subsection{Natal kick models}

\citet{Hobbs0507} performed a study of 233 galactic pulsars proper motions and find out that for young (i.e. with characteristic age $\tau_{\rm c}=P/2\dot{P}<3\yr$) pulsars the velocity distribution is well fitted with Maxwellian distribution with $\sigma=265$ km/s. Velocities of such a population may quite well resemble the NKs of NSs. However, it is not obvious, if the NKs of BHs follow the same distribution. Especially, if NKs are driven by the asymmetries in ejecta, the post-SN fall-back may significantly lower the BH's velocity. Proper motion measurements are available only for a few BHs residing in XRBs \citep{MillerJones1403} and the connection between the current peculiar velocity and the NK is not straightforward \citep[e.g.][]{Fragos0906}. On the other hand, the analysis of the positions of XRBs harbouring BHs may suggest that distributions of NKs of BHs and NSs are similar \citep[e.g.,][however, see \citeauthor{Mandel1602} \citeyear{Mandel1602}]{Jonker0410,Repetto1705}. 

In the standard model, we lower the remnant's NK proportionally to the fall-back of material after a SN explosion. Precisely, the NK for a compact object is equal

\begin{equation}\label{eq:NK_H}
	V_{\rm kick}=V_{\rm kick, Maxwell} (1-f_{\rm FB}),
\end{equation}

\noindent where $V_{\rm kick, Maxwell}$ has a Maxwellian distribution with $\sigma=265$ km/s and $f_{\rm FB}$ indicates what fraction of ejected mass falls back onto a compact object \citep[see ][]{Fryer1204}. This model produces both low-velocity and high-velocity BHs which is in agreement with observational estimates for XRBs \citep[e.g.][]{Belczynski1603}.

If NKs are driven by neutrino based mechanism, it is predicted that the imposed momentum will be the same for BHs and NSs \citep[e.g.,][]{Janka1309,Rodriguez1611}. Such NKs will be inversely proportional to the mass of a BH ($V_{\rm kick} \approx M_{\rm BH}^{-1}$). In \nkr\ model we assume that NSs' NK distribution has a Maxwellian shape with $\sigma=265$ km/s, but BHs
NK distribution is lowered. Precisely, 

\begin{equation}\label{eq:NK_R}
	\begin{array}{lcl}
		V_{\rm kick,NS}&=&V_{\rm kick, Maxwell},\\
		V_{\rm kick,BH}&=&V_{\rm kick, Maxwell} \frac{M_\mathrm{max,NS}}{M_{\rm BH}},
	\end{array}
\end{equation}

\noindent where $M_\mathrm{max,NS}=2.5\msun$ is an adopted limit for a NS mass above which it collapses to a BH. It must be noted that, even for low-mass BHs ($\sim5$--$7\msun$), such a prescription will produce significantly lower NKs, than the NKs of NSs, what stands in contradiction with the results of \citet[e.g.][]{Repetto1705}. Nevertheless, we included this model as a parameters study.

Finally, we incorporated a simple model proposed by \citet{Bray1811}, who suggested that the NK distribution can be described as a linear function of a ratio between the mass of the ejecta ($M_{\rm ejecta}$) and the mass of the remnant ($M_{\rm remnant}$). We applied (model \nkbe) their fit to the observations provided by \citet{Hobbs0507} according to which the NK in 3D is given by

\begin{equation}\label{eq:NK_BE}
	V_{\rm kick}=60\,\frac{\mathrm{km}}{\rm s} \left(\frac{M_{\rm ejecta}}{M_{\rm NS/BH}}\right) + 130\,\frac{\mathrm{km}}{\rm s},
\end{equation}
\noindent where $M_{\rm ejecta}$ is mass ejected from the star during SN and $M_{\rm NS/BH}$ is mass of a resulting NS, or a BH. Although \citet{Bray1811} provided their fit only for NSs, we apply it also to BHs as a parameter study.

\subsection{Mergers of binary components}\label{sec:methods_merger}

Stellar mergers may constitute a significant fraction of massive stars \citep[e.g.][]{Langer1209}, and therefore, progenitors of BHs. Head-on collisions may play a significant role in dense stellar systems like globular clusters \citep[e.g.][]{Glebbeek1310}, however, stellar mergers in field populations are more probable due to orbital angular momentum loss (e.g. during CE phase). \citet{Podsiadlowski9205} predicted that $\sim10\%$ of $8<M<20\msun$ primaries merge with their companions before a SN and \citet{deMink1402} showed that $\sim8^{+9}_{-4}\%$ of massive single stars may actually be products of binary mergers. Objects like Red Novae, or Blue Stragglers are thought to be observed during, or after the merging process \citep[e.g.,][]{Blagorodnova1701,Leonard8907,Kochanek1409}.
 
In case of non-DCO mergers (including mergers of a compact object with a non-compact object), a merger process is not well understood. Especially, the amount of mass ejected from the system, which is an important factor for post-merger evolution, is poorly constrained. A product of a merger of two MS stars may evolve similarly to a star which was always single, but if evolved stars are involved, the post-merger evolution is more complicated \citep{Glebbeek1310}. Especially, a merger outcome may evolve unlike any single star \citep{VignaGomez1903}.

We assume that non-DCO mergers occur in the following situations:
\begin{itemize}
    \item failed envelope ejection during a CE event \citep[e.g.][]{Justham1412},
    \item donor in a CE phase haven't developed a clear boundary between core and the envelope \citep[e.g. MS and HG donors;][]{Ivanova0402}, 
    \item donor's radius exceeds two times its Roche lobe radius 
($R_{\rm donor}>2R_{\rm donor, RL}$) during RLOF.
\end{itemize}
\noindent Due to the fact that the merger physics and post-merger evolution is poorly understood, we provide information on the binary parameters just before the merger to allow for different approaches to this conundrum.

As far as DCO mergers are concerned, many studies were devoted to an in-depth analysis of this phenomenon in stellar populations \citep[e.g.,][]{Lipunov9705,Dominik1211,Dominik1312,Dominik1506,Mennekens1404,Mandel1605,Belczynski1606}. DCO mergers affect the BH populations, especially the BH mass distribution. As a part of our results, in the public database we provide information about DCO formation and estimate the time to merger using the formula of \citet{Peters6411}. In order to give predictions and estimate the fraction of BHs which come from DCO mergers, we assumed that the mass loss during a DCO merger is negligible.

In this work, in order to estimate the population of SBH originating from binary mergers, we adopt a simplistic approach of Olejak et al. (in prep.). We include only main channels responsible for $>95\%$ of all the mergers. Specifically, for different evolutionary types of stars we use the following procedure:

\begin{itemize}
    \item MS+MS - Outcome is a MS star. We assume that half of the mass of the lighter component is lost in the process.
    \item MS+HeS - Outcome is a helium star (HeS) star. We assume that half of the mass of the MS component is lost in the process.
    \item HeS+HeS - Outcome is a HeS star. We assume that half of the mass of the lighter component is lost in the process.
    \item NS+MS/HeS - We assume that half of the mass of the MS/HeS star is lost in the process and that it becomes a NS/BH if its final mass is lower/higher than $M_{\rm max,NS}=2.5\msun$.
    \item BH+MS/HeS - We assume that half of the mass of the MS/HeS star is lost in the process and that it becomes a BH.
    \item BH+NS/BH - The outcome is a BH with a mass equal to the total mass of the binary before the merger.
\end{itemize}

We neglect other types of mergers as we found that they constitute only a small fraction ($\lesssim5\%$) of all mergers and predictions for their outcomes are even less certain. For post-merger MS and HeSs we calculate their further evolution assuming that they are on ZAMS, or zero-age helium main-sequence, respectively, in order to find BH predecessors and calculate compact object's mass. We note that presented approach is only quantitative, but helps to estimate the importance of stellar merger for BH populations.

\subsection{General simulation properties}

In this work, we concentrate on a description of BH populations originating from massive binaries, which we define as binaries with a primary's (i.e., heavier star on ZAMS) initial mass $\mzamsa>10\msun$. Our analysis does not include BHs in triples and higher-order systems \citep[e.g.][]{Antonini1706}, as well as BHs in binaries formed due to stellar encounters in dense stellar systems \citep[e.g.][]{Banerjee1705}.

The total simulated stellar mass is $4.8\times10^8\msun$ for the \std\ model and $1.1\times10^9\msun$ ($2.4\times10^8\msun$) for the \imfs (\imff) model. Throughout this paper, we assume $50\%$ binary fraction on ZAMS for low-mass stars ($M_{\rm ZAMS}<10\msun$) and $100\%$ binary fraction for heavier stars ($M_{\rm ZAMS}\geq10\msun$). The simulated stellar mass corresponds to about $\sim0.8\%$ ($1.9\%$ for \imfs\ model, or $0.4\%$ for \imff\ model) of the stellar mass of the MW equivalent galaxy \citep[MWEG; $M_{\rm MW}\approx6\times10^{10}\msun$; e.g.][]{Licquia1506}.

For the purpose of presentation, all results are scaled to the stellar mass of a MWEG \citep[$M_{\rm MW}\approx6\times10^{10}\msun$;][]{Licquia1506} and the SFH is chosen to be constant through the last $10\gyr$ (${\rm SFR}=6\,M_{\odot}\,yr^{-1}$). Such a model, although simple, allows us to draw general conclusions. Raw results may be scaled by using any realistic total stellar mass and any SFH may be applied. In general, the changes in SFH are not as important as the total mass of formed stars, as there is only a small delay time between the formation of a binary (ZAMS) and formation of a BH (typically less than $a\; few \times10\myr$), which are, typically, single. We note, that for a part of the BH population the SFH may, actually, be important, e.g. for DCOs, which have steep time-to-merger distribution \citep[$t_{\rm merge}\propto t^{-1}$; e.g.][]{Dominik1211}. Additionally, if the spacial distribution of BHs is concerned, the change of SFH may play a role, because it affects the time which BHs have to spread throughout the space. What is more, BHs formed earlier have more time to be ejected from the stellar systems due to their NKs, or as a result of dynamical interactions. Deeper analysis of changes in spacial distributions, SFH, and chemical evolution are left for separate studies (e.g. Olejak et al. in prep.).

\section{Results} 

\begin{deluxetable*}{lcrlrlrlrl}
    \tablewidth{\textwidth}
    \tablecaption{Number of BHs in the Milky Way Equivalent Galaxy}
    \tablehead{ Model & $N_{\rm BH,tot}$ & \multicolumn{2}{c}{$N_{\rm dSBH}$} & \multicolumn{2}{c}{$N_{\rm BHB}$} & \multicolumn{2}{c}{$N_{\rm BH,DCO}$} & \multicolumn{2}{c}{$N_{\rm mSBH}$}}
    \startdata
    \std  & \sci{1.1}{8} & \sci{5.4}{7} & ($49.7\%$) & \sci{1.5}{6} & ($1.4\%$) & \sci{5.1}{6} & ($4.7\%$) & \sci{4.8}{7} & ($44.2\%$) \\
    \midz & \sci{1.1}{8} & \sci{3.4}{7} & ($30.6\%$) & \sci{2.0}{6} & ($1.8\%$) & \sci{1.2}{7} & ($10.8\%$) & \sci{6.3}{7} & ($56.8\%$) \\
    \lowz & \sci{1.2}{8} & \sci{3.7}{7} & ($30.0\%$) & \sci{1.5}{6} & ($1.2\%$) & \sci{1.9}{7} & ($15.4\%$) & \sci{6.6}{7} & ($53.4\%$) \\
    \ssa  & \sci{1.1}{8} & \sci{6.6}{7} & ($58.8\%$) & \sci{3.4}{6} & ($3.0\%$) & \sci{6.9}{6} & ($6.1\%$) & \sci{3.6}{7} & ($32.1\%$) \\
    \nkr  & \sci{1.1}{8} & \sci{6.4}{7} & ($58.1\%$) & \sci{5.9}{4} & ($0.1\%$) & \sci{1.1}{4} & ($0.0\%$) & \sci{4.6}{7} & ($41.8\%$) \\
    \nkbe & \sci{1.1}{8} & \sci{6.5}{7} & ($58.5\%$) & \sci{5.4}{4} & ($0.0\%$) & \sci{2.7}{3} & ($0.0\%$) & \sci{4.6}{7} & ($41.4\%$) \\
    \imff & \sci{2.7}{8} & \sci{1.4}{8} & ($52.6\%$) & \sci{3.2}{6} & ($1.2\%$) & \sci{1.3}{7} & ($4.9\%$) & \sci{1.1}{8} & ($41.3\%$) \\
    \imfs & \sci{3.4}{7} & \sci{1.6}{7} & ($46.8\%$) & \sci{5.6}{5} & ($1.6\%$) & \sci{1.6}{6} & ($4.7\%$) & \sci{1.6}{7} & ($46.8\%$) \\
    \enddata
    \tablecomments{Number of BHs for different tested models (see Tab.~\ref{tab:models}). Values are presented for a simple MW model with a total simulated stellar mass of $6\times10^{10}\msun$ and constant star formation during the last $10\gyr$. Column headers stand for: $N_{\rm BH,tot}$ - total estimated number of BHs; $N_{\rm dSBH}$ - number of single BHs from disrupted binaries; $N_{\rm BHB}$ number of binaries harbouring a BH and a non-compact companion; $N_{\rm BH,DCO}$ - number of BHs in DCOs (BH+BH is counted as two BHs; see Tab.~\ref{tab:dco}); $M_{\rm mSBH}$ - a rough estimate of the number of stellar mergers (including DCO) which will be massive enough to form a BH (see Sec.~\ref{sec:methods_merger}).} 
    \label{tab:results}
\end{deluxetable*}

The main results of the simulations are summarized in Tab.~\ref{tab:results}. Most of the BHs are predicted to exist as SBHs either as a result of a binary disruption (dSBH), or a stellar merger (mSBH; with a BH involved, or producing a star massive enough to form a BH). The population of BHs in binaries (both in DCOs and in BHBs) is about an order of magnitude smaller and in the case of models with increased BH NKs (\nkr\ and \nkbe) even two orders of magnitude smaller. Only a very small fraction of the BHBs are the interacting ones (see Sec.~\ref{sec:mt}).

\begin{figure}[t]
    \centering
    \includegraphics[width=1.0\columnwidth]{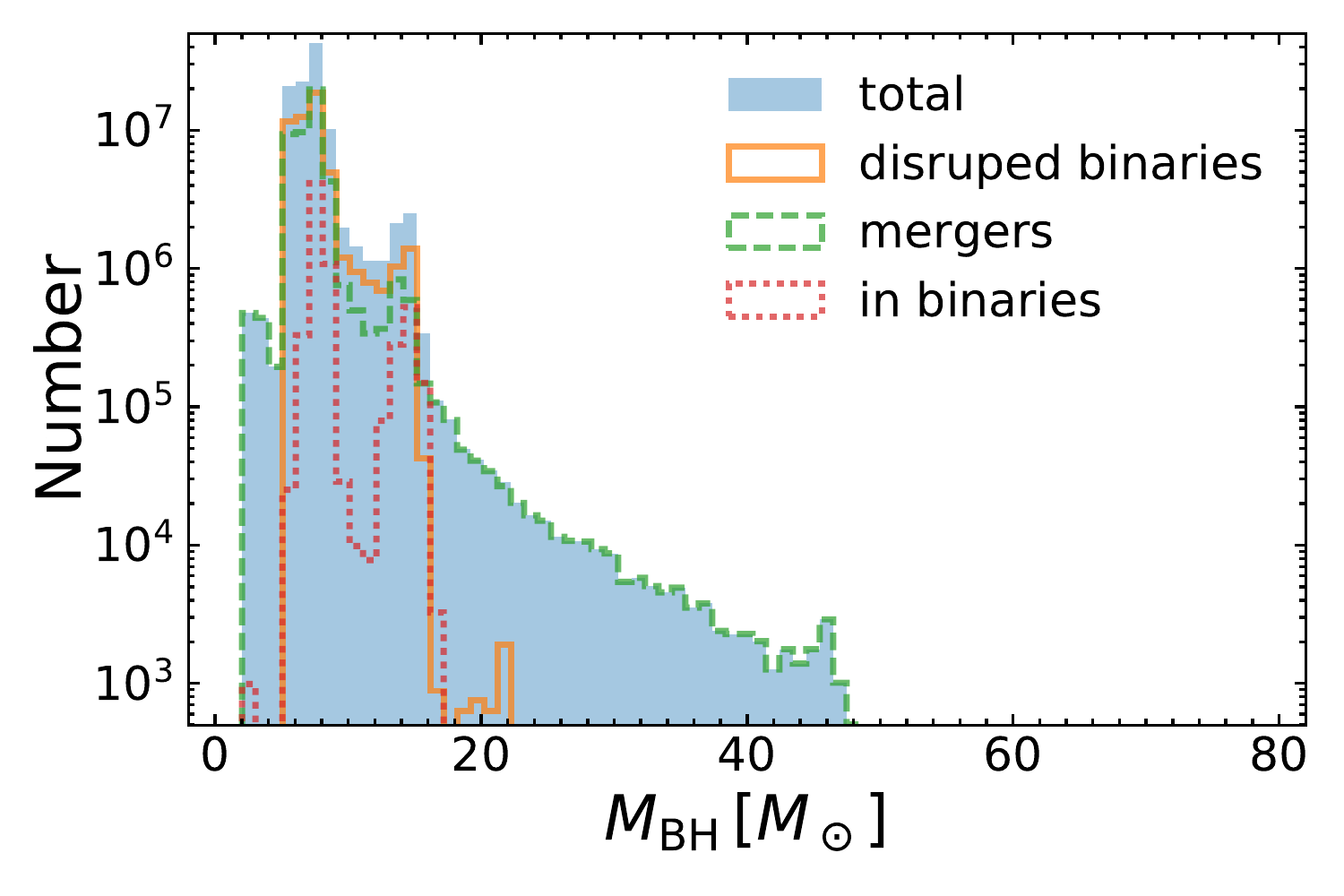}
    \caption{The BH mass distribution for the \std\ model containing single BH from disrupted binaries and mergers, and BHs residing in binaries. The results are scaled for the MWEG with total stellar mass of $6\times10^{10} \msun$ and constant SFR throughout the last $10 \gyr$.}
    \label{fig:mass}
\end{figure}

The initial separation influences significantly the fate of a massive binary. In general picture, if the separation is very large ($a\gtrsim 3,000\rsun$) the binary will mostly disrupt and produce SBHs (Sec.~\ref{sec:SBH}). A low separation ($\azams\lesssim 30\rsun$) will lead frequently to a non-DCO merger (Sec.~\ref{sec:mergers}), which may be massive enough to form a SBH. Objects with medium initial separations have much more uncertain fate and may merge, become disrupted, or form a bound binary with at least one BH inside (Sec.~\ref{sec:bhb}). In the case of a binary harbouring a BH, the system after some time may still merge (e.g. as a double compact object merger, Sec.~\ref{sec:dco}), or be disrupted during the formation of a second compact object (e.g., \rSBHb\ formation route, Sec.~\ref{sec:SBH_er}). 

\begin{deluxetable}{ll}
    \tablewidth{\columnwidth}
    \tablecaption{Typical formation routes of BHs}
    \tablehead{ Route & evolutionary route }
    \startdata
    \multicolumn{2}{c}{Single BHs from disrupted binaries}\\
    \rSBHa & MT1(1/2/4/5-1/2/4/5) SN1 Disruption SN2 \\
    \rSBHb & SN1 SN2 Disruption\\\\
    \multicolumn{2}{c}{BH XRBs}\\
    \rlmxba & CE1(4/5-1;7/8-1) SN1 MT2(14-1/2/3/4/5/6) \\
    \rlmxbb & CE1(4/5-1;7/8-1) SN1 MT2(13-1/2/3) AICBH1\\
     & \hfill MT2(14-1/2/3)\\\\
    \multicolumn{2}{c}{DCOs}\\
    \rdcoa & SN1 SN2\\
    \rdcob & MT1(1/2/4/5-1/2/4) SN1\\
     & \hfill CE2(13/14-4/5;13/14-7/8) SN2\\\\
    \multicolumn{2}{c}{Mergers}\\
    \rmer & CE1(2/3/4-1;7-1) MT1(7-1) Merger\\
    \enddata
    \tablecomments{Schematic representations of the evolutionary routes for typical BH formation channels. Wide BHBs (Sec.~\ref{sec:bhb_wide}) are not included in the table as their evolution does not include any interactions and their evolutionary route is 'SN1' by definition. Details of the post-merger evolution are not included in our study. Symbols in evolutionary routes represent: SN1/2 - supernova of the primary/secondary; MT1/2 - mass transfer (primary/secondary is a donor); CE1/2 - common envelope (primary/secondary is a donor; numbers in parenthesis represent typical primary evolutionary type (left) and secondary (right); first two numbers represent initial types (prior to CE), whereas, last two numbers represent the final types (after CE)); AICBH1 - accretion induced collapse of a NS primary to a BH (assumed to occur after a NS obtain the mass of $M\geq M_\mathrm{max,NS}=2.5\msun$). Evolutionary types (numbers inside parenthesis) represent: 1-main sequence, 2-Hertzsprung gap, 3-red giant, 4-core helium burning, 5-early asymptotic giant branch, 6-thermal pulsing asymptotic giant branch, 7-helium main sequence, 8-evolved helium star, 13-neutron star, 14-black hole.}
    \label{tab:er}
\end{deluxetable}

\subsection{Single BHs from disrupted binaries}\label{sec:SBH}

If a system hosting a BH progenitor becomes disrupted, what may occur before, after, or during the BH formation, a compact object becomes an dSBH (we use name dSBH to represent a SBH from a disrupted binary, as SBH can also form due to mergers; see Sec.~\ref{sec:mergers}). Binaries may become disrupted due to a NK which one component obtains after a SN explosion, due to a Blaauw kick \citep[e.g. because of a significant loss of mass ($\gtrsim50\%\;M_{\rm tot}$) from a binary in SN explosion;][]{Blaauw6105}, or interaction with a third star. Although in our simulations we included only isolated binaries, we note that interactions, even in sparse stellar systems like the Galactic disk, may significantly alter the binary evolution \citep{Kaib1402,Klencki1708}.

\begin{deluxetable}{lccccl}
    \tablewidth{\columnwidth}
    \tablecaption{Single BHs from disrupted binaries}
    \tablehead{ Model & Route & Number & $a_{\rm ZAMS}[\rsun]$  }
    \startdata
    \std & \rSBHa & $3.6\times10^7 (67\%)$ & $28\div400$ \\
     & \rSBHb & $1.7\times10^7 (32\%)$ & $>4400$ \\
    \midz & \rSBHa & $2.0\times10^7 (58\%)$ & $20\div800$ \\
     & \rSBHb & $1.2\times10^7 (33\%)$ & $>7700$ \\
    \lowz & \rSBHa & $1.5\times10^7 (40\%)$ & $17\div180$ \\
     & \rSBHb & $2.0\times10^7 (55\%)$ & $>1600$ \\
    \ssa & \rSBHa & $4.0\times10^7 (61\%)$ & $28\div1000$ \\
     & \rSBHb & $2.4\times10^7 (36\%)$ & $>6200$ \\
    \nkr & \rSBHa & $4.0\times10^7 (62\%)$ & $28\div400$ \\
     & \rSBHb & $2.2\times10^7 (35\%)$ & $>5200$ \\
    \nkbe & \rSBHa & $4.1\times10^7 (63\%)$ & $28\div400$ \\
     & \rSBHb & $2.2\times10^7 (34\%)$ & $>5200$ \\
    %\ssb & \rSBHa & $2.5\times10^7 (53\%)$ & $28\div400$ \\
    % & \rSBHb & $1.8\times10^7 (38\%)$ & $>3100$ \\
    \imff & \rSBHa & $9.4\times10^7 (66\%)$ & $28\div430$ \\
     & \rSBHb & $4.7\times10^7 (33\%)$ & $>5400$ \\
    \imfs & \rSBHa & $1.1\times10^7 (68\%)$ & $28\div400$ \\
     & \rSBHb & $5.1\times10^6 (31\%)$ & $>5000$ \\
    \enddata
    \tablecomments{Number of single BHs originated from disrupted binaries. Only main evolutionary routes are included). The initial separation ($a_\mathrm{ZAMS}$) is the main differing factor between the main routes. See Sec.~\ref{sec:SBH_er} for details.}
\label{tab:SBH}
\end{deluxetable}

We found out that the tested initial parameters distributions have little effect on the properties of the resulting population of dSBHs. The steepness of the IMF changes number of resulting dSBH by $\Delta N_{\rm dSBH}\approx70\%/160\%$ due to the decreased/increased relative number of BH progenitors for \imfs/\imff\ models. The effect of metallicity and NK models is not as strong ($\Delta N_{\rm dSBH}\approx40\%$ and $20\%$, respectively).

\subsubsection{Typical evolutionary routes}\label{sec:SBH_er}

We found that two evolutionary routes (see Tab.~\ref{tab:er}) dominate most of the dSBH production and are common for all models. Number of dSBHs produced through these routes and for different models is presented in Tab.~\ref{tab:SBH}. Both channels involve a binary initially composed of two massive stars that will undergo SN explosions, or collapse directly into BHs. The former case may lead to a binary disruption (due to NK and/or mass loss). The main difference between the routes comes from the initial separation. In the case of \rSBHa, the initial distance between stars is lower than $\sim1000\rsun$, whereas for \rSBHb\ it is usually larger then $\sim5000\rsun$. Consequently, the former route leads through an interaction phase (typically the MT), whereas a binary evolving through the latter one experiences no interactions.

\begin{figure}[t]
    \centering
    {\large\rSBHa}\\
    \includegraphics[width=1.0\columnwidth]{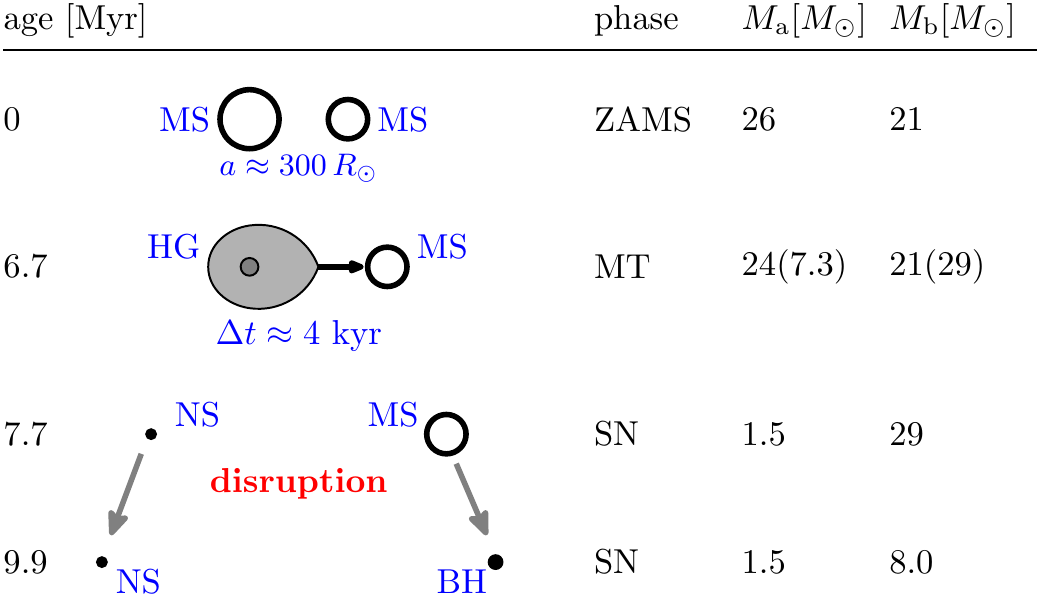}\\
    \vspace{0.5cm}
    {\large\rSBHb}\\
    \includegraphics[width=1.0\columnwidth]{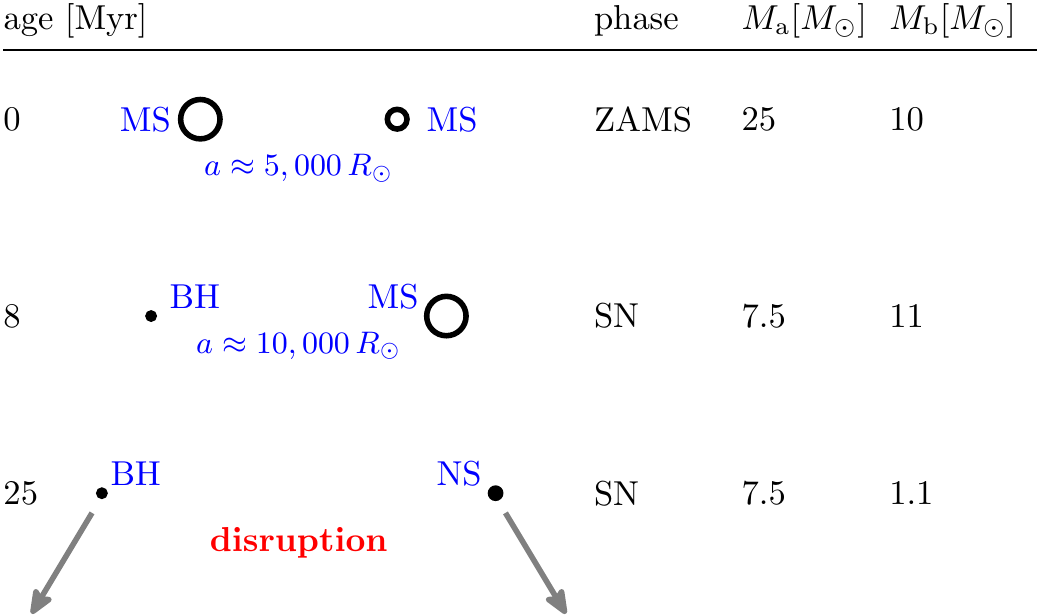}
    \caption{Schematic representations of the typical evolutionary scenarios leading to the formation of a single BHs originating from disrupted binaries (dSBH; Sec.~\ref{sec:SBH_er}). 'Age' represents the time since ZAMS, \ma/\mb\ stands for the mass of the primary/secondary. The evolutionary phases appearing in the figure are the following: ZAMS - zero age main sequence; MT - mass transfer; SN - supernova, i.e. compact object formation. The highlighted stages in stellar evolution: MS - main sequence; HG - Hertzsprung gap; NS - neutron star; BH - black hole. The numbers in parentheses represent the masses at the end of the MT phase.}
    \label{fig:rdSBH}
\end{figure}

For a typical binary evolving through \rSBHa\ route (Fig.~\ref{fig:rdSBH}, upper plot), the primary is about $26\msun$ on ZAMS, whereas secondary is slightly lighter ($\sim21\msun$). The separation is modest ($\sim300\rsun$). The primary evolves faster and after $\sim6.7\myr$, while expanding as a HG star, fills its Roche lobe and a MT begins onto the secondary, which is still on its MS. Although the MT is relatively short ($\sim4\kyr$), the primary loses its hydrogen envelope and becomes a $\sim7\msun$ helium star. Half of the expelled envelope is accreted by the secondary which grows to $\sim29\msun$ still being on its MS. After about $1\myr$ primary goes through SNIb/c and becomes a $\sim1.5\msun$ NS. A NK disrupts the system. Due to its significant mass, the velocity which the secondary obtains is very low ($\sim5\kms$). The massive secondary needs only $\sim2\myr$ to become a BH with a mass of $\sim8\msun$. The NK in a case of such a BH is negligible, therefore, its velocity doesn't change significantly. It is noteworthy that most of the BHs formed through \rSBHa\ route originate from secondary stars, i.e. less massive ones on ZAMS.

The \rSBHa\ channel is similar to the one proposed for the formation of single massive stars (potential BH progenitors) by \citet{Renzo1904}. Using different population synthesis code, they found that $86^{+11}_{-9}\%$ (depending on the model parameters) of binaries evolving through this channel will become disrupted during the first SN.

In the other typical route (\rSBHb; Fig.~\ref{fig:rdSBH}, bottom plot) the primary is about $25\msun$ on ZAMS, whereas the secondary is $\sim10\msun$. The initial separation is as large as $\sim6000\rsun$. As a consequence, their Roche lobes are huge and no interaction is possible throughout their evolution. The heavier star evolves faster and in $\sim8\myr$ forms a $\sim7.5\msun$ BH with a small NK. The separation grows to $\sim10,000\rsun$ what results from a loss of mass in stellar wind from both stars. After additional $\sim17\myr$ secondary explodes as a SN and forms a NS. A strong NK disrupts the system. Due to a large mass ratio between compact objects ($q\approx7$) the post-disruption velocity of the BH is relatively small ($\sim5\kms$).

\subsubsection{Mass distribution of dSBHs}\label{sec:SBH_mass}

\begin{figure}[t]
    \centering
    \includegraphics[width=1.0\columnwidth]{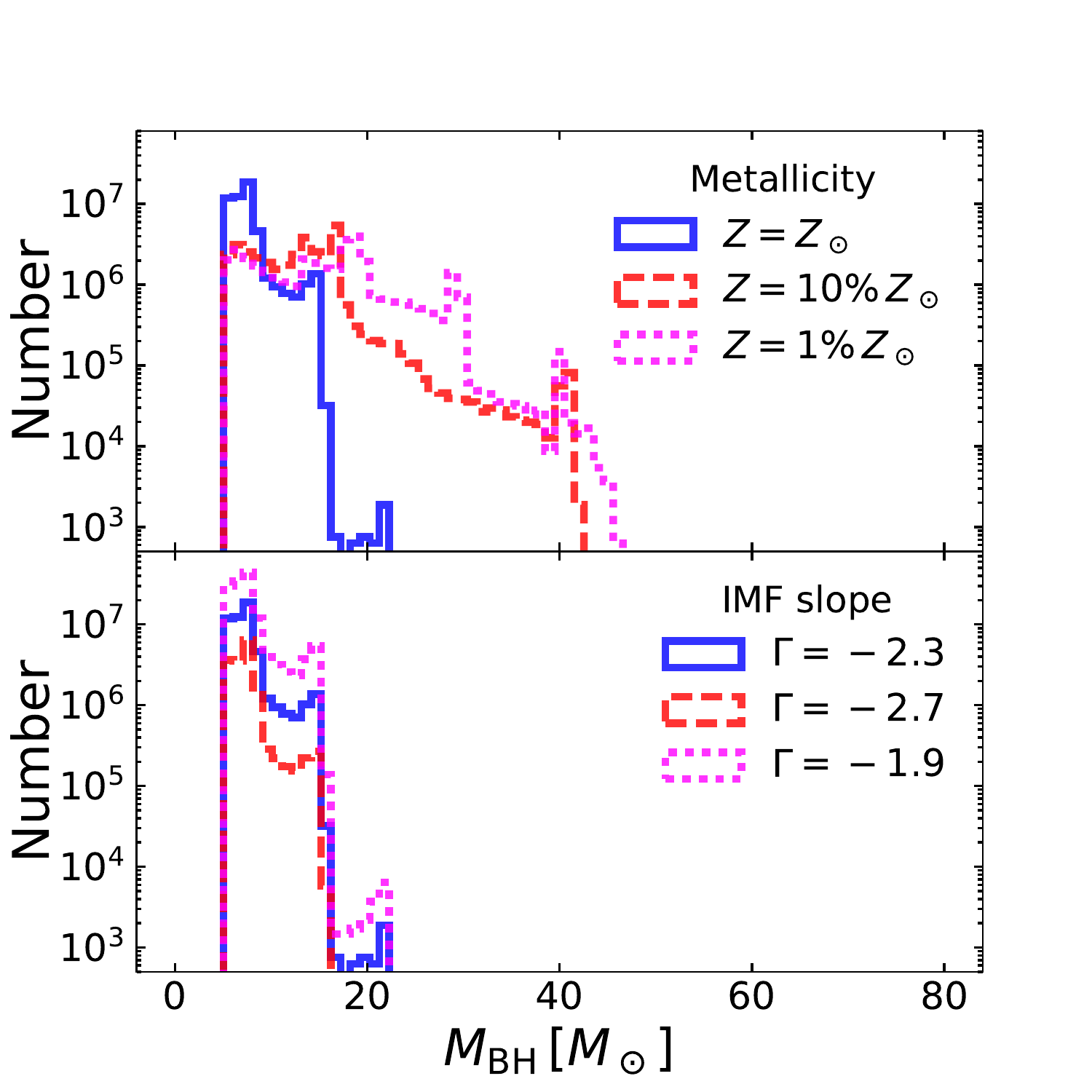}
    \caption{Mass distribution of the single BHs from disrupted binaries. The location of the main peaks depends on metallicity (upper plot; e.g., $\sim7$--$8\msun$ and $\sim15\msun$ for \std\ model, and $\sim15$--$30\msun$ and $\sim40\msun$ for \midz\ and \lowz\ models). We note, that the peak at $\sim40\msun$ at \midz\ model results from mass loss in stellar winds, whereas in \lowz\ model it is a consequence of pair pulsation SNe. Two main evolutionary routes, \rSBHa\ and \rSBHb, have similar mass distributions with the peaks located at the same masses for all models with the same metallicity. All distributions reveal a high-mass extension (a peak at $\sim22\msun$ for \std\ model and a tail between $\sim40$--$45\msun$ for models \midz\ and \lowz), which is a result of the mass accretion onto a BH, or its progenitor, prior to disruption.}
    \label{fig:dSBH_mass}
\end{figure}

The dSBHs' mass distribution is presented in Fig.~\ref{fig:dSBH_mass}. The shape is similar to the distribution of BH masses in single star evolution \citep[compare e.g. $M_\mathrm{remnant}(M_\mathrm{ZAMS})$ relations in][]{Belczynski1005}. The main peak ($\sim7$--$8\msun$ for solar metallicity (\std\ model; $Z=\zsun$), or $\sim15$--$30\msun$ for lower metallicities (models \midz\ and \lowz; $Z=10\%\zsun$ and $Z=1\%\zsun$, respectively)) relates to BHs formed from $\sim20$--$35\msun$ progenitors through failed SN explosion \citep[e.g.][]{Fryer1204,Belczynski1209}. The second peak ($\sim15\msun$ for \std, or $\sim40\msun$ for \midz\ model) comes from BHs originating from most massive stars ($\gtrsim100\msun$) on ZAMS, which lose a large part of their mass in stellar wind. In the case of low metallicity environments (model \lowz) the peak results from the pair-pulsation SN \citep[e.g.][]{Belczynski1610}, which prevents the formation of heavier BHs in single star evolution. The third peak ($\sim22\msun$ for \std, or the tail extending to $\sim45\msun$ for \midz\ and \lowz) is formed through binary interactions and is not present in distributions for single star evolution \citep[compare][]{Belczynski1005,Belczynski1610}.

The similarity of the mass distributions of BHs for single stars and binaries is a direct consequence of the fact that massive binaries, which later on became disrupted, usually interact only little (mass of BH progenitor is not changed significantly; \rSBHa), or not at all (\rSBHb), so in most cases the binary components evolve as in isolation. 

A notable difference is the presence of the high-mass peak and tail in the mass distributions. It is most prominent for the models with solar metallicity where the tail extends up to $\sim22\msun$, whereas the heaviest BHs forming at that metallicity from single stars are only $\sim15\msun$ \citep{Belczynski1005}. These massive SBHs constitute only a tiny fraction of the population ($<1\%$). Typically, the initial total mass of a system in which such over-massive BHs are formed is above $200\msun$ and a mass ratio is $\sim1$. The primary, being slightly heavier, is the first to expand and fill its Roche lobe while being on a HG. At the moment of interaction both stars have masses of about $60\msun$ due to strong mass losses in stellar wind. When the mass transfer ceases, the primary becomes a $\sim25\msun$ core helium burning (CHeB) star. The secondary being now much heavier ($\sim80\msun$), appears rejuvenated \citep[e.g.][]{Tout9711}, evolves much faster and after $\sim400\kyr$ forms a BH first with a mass of $\sim21\msun$. The primary needs additional $100\kyr$ to become a $7.3\msun$ BH. During the second SN, NK easily disrupts a wide ($\sim15,000\rsun$) binary.

\subsubsection{Significance of metallicity}

\begin{figure}[t]
    \centering
    \includegraphics[width=1.0\columnwidth]{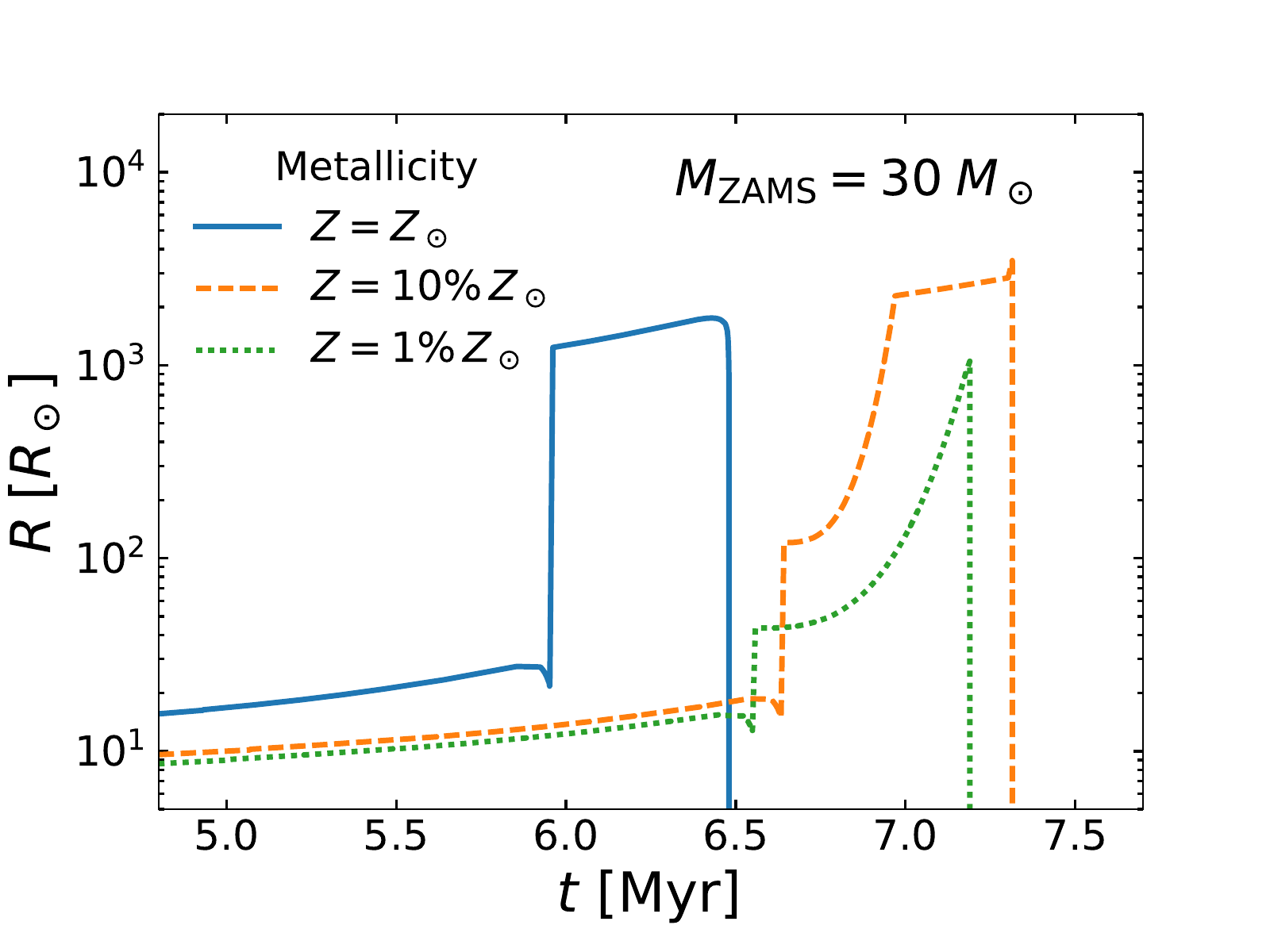}
    \caption{Radius evolution of a star with a ZAMS mass of $M_{\rm ZAMS}=30\msun$ for three different metallicites. The largest radii are obtained for moderate metallicity ($Z=10\%\zsun$; \midz\ model), whereas the smallest for low metallicity ($Z=1\%\zsun$; \lowz\ model).}
    \label{fig:exp-Z}
\end{figure}

The relation between the number of dSBH and metallcity was found to be non-monotonic. The largest number of dSBHs in respect to metallicity is produced in the \std\ model ($5.4\times10^7$), whereas the smallest is in the \midz\ model ($3.4\times10^7$). For \lowz\ model the number of dSBHs is slightly higher ($3.7\times10^7$) than in the \midz\ model.  

The number of dSBHs produced through the \rSBHa\ route increases monotonically with metallicity. The higher is the metallicity, the stronger is the line-driven stellar wind. Only a fraction of this mass may be accreted by a companion, thus bulk of it leaves the system what results in orbital expansion. Consequently, the stronger is the stellar wind, the wider the binary becomes. Simultaneously, the stronger mass loss also makes the companion lighter during the NS formation. Wider orbits and lighter companions make the system easier to disrupt, what as a result, produces more SBHs. 

A different mechanism results in the non-monotonic relation between the metallicity and the number of BHs produced through the \rSBHb\ route. Metallicity affects the evolutionary expansion of the stars which happens to be the strongest in \midz\ environment and the smallest in the \lowz\ environment (Fig.~\ref{fig:exp-Z}). Only binaries with separations high enough to avoid Roche lobe filling may evolve without interactions. The number of such systems will be smaller if the nuclear expansion is on average higher as in the case of \midz\ model.

\subsubsection{Initial BH velocities and high velocity dSBHs}\label{sec:highv}\label{sec:SBH_nk}

When a SBH forms (after disruption of the binary, or BH formation, whatever happens later) its velocity is not only a result of the motion in the gravitational potential (e.g. of a galaxy), but also preceding evolutionary processes, like NK, Blaauw kick, and binary disruption. The velocity which is the result of the latter processes calculated at the moment of the binary disruption, or the formation of the BH (whatever happens later) we call here an initial BH's velocity (\ivbh). Such a definition in which we do not involve the motion in the gravitation potential allows for application of our results to any gravitational potential (e.g. different galaxies, or different models of the MW galaxy). In this work, we provide the magnitudes of 3D velocities (i.e. lengths of the velocity vectors) and assume that the distribution of NKs is isotropic. We note that in a realistic situation, the velocity of a dSBH will be modified by the gravitational potential of a stellar system and interactions with other stars. Although these effects may also affect the binary evolution, the short dSBH formation timescale ($\lesssim50\myr$) allow us to assume that the effect is usually negligible.

The highest values of \ivbh\ are obtained for BHs with the highest NKs. We found that majority of dSBHs with the highest initial velocities ($\ivbh>300$ km/s) within the highest $10\%$ were formed through \rSBHa\ route and involve lowest BH masses and the shortest pre-disruption periods. We note, that \citet{Renzo1904} found that the majority of the secondaries ($\sim90\%$) after the disruption will have low velocities ($\lesssim20$ km/s) what agrees with our predictions for standard NK model.

\begin{figure}[t]
    \centering
    \includegraphics[width=1.0\columnwidth]{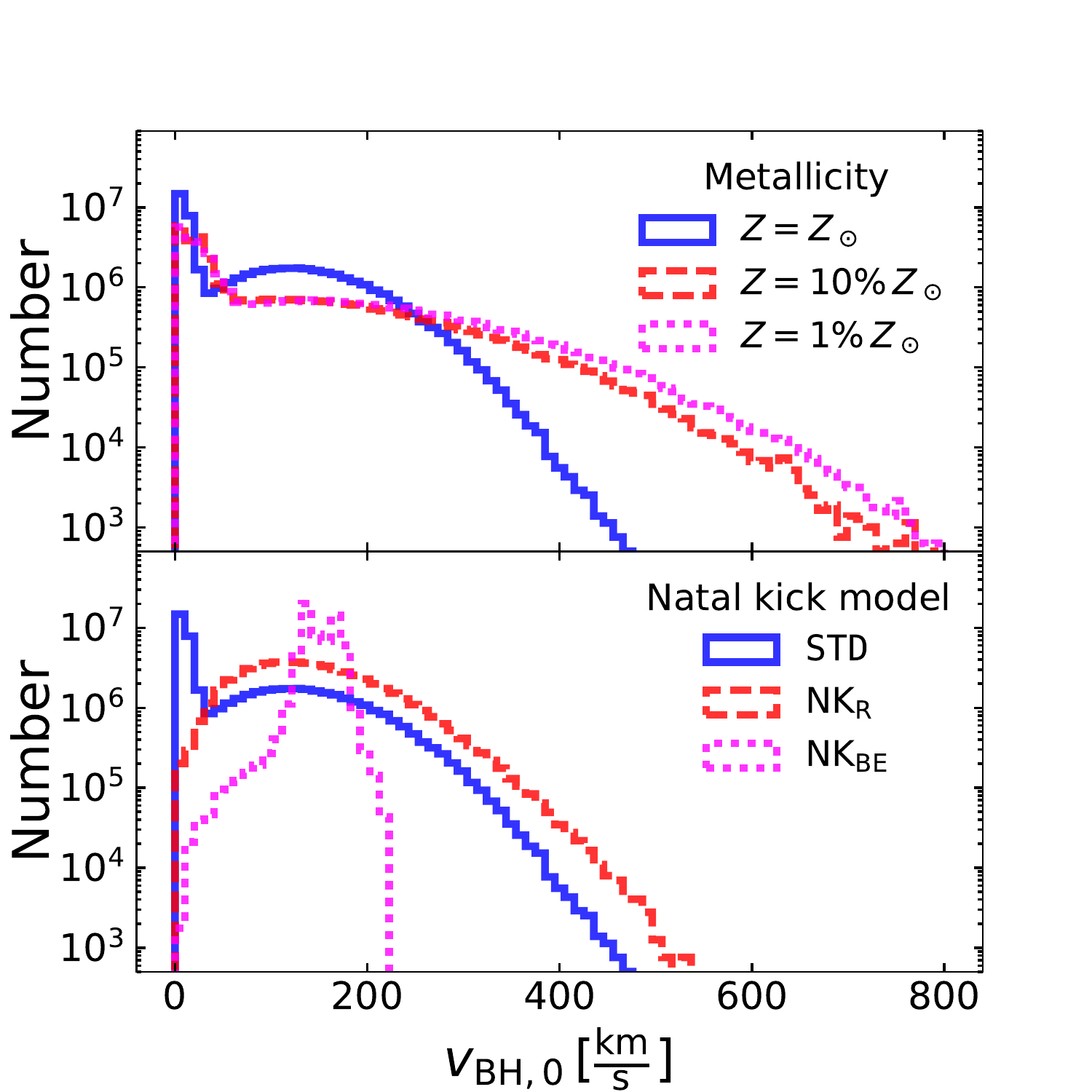}
    \caption{Distributions of initial velocities ($\ivbh$) of single BHs originating from disrupted binaries. All models with standard \citep{Hobbs0507} NK prescription have a similar bimodal distribution. \nkr\ and \nkbe\ NK prescriptions give significantly different distributions. NKs become larger for lower metallicities what results in wider initial velocity distributions. We note, that local escape velocity from the Milky Way (at the radius of $\sim8.3$ kpc) is $\sim520$ km s$^{-1}$ and drops to $\sim380$ km s$^{-1}$ at the radius of 50 kpc \citep[e.g.][]{Williams1706}.}
    \label{fig:dSBH_kick}
\end{figure}

\ivbh\ distributions for all tested NK models are presented in Fig.~\ref{fig:dSBH_kick}. The maximal \ivbh\ of a dSBH may reach $\sim500\kms$ for \std\ model and nearly $\sim1000\kms$ for \lowz\ model. The higher velocities for lower metallicities are the result of a formation of closer binaries through \rSBHa\ route, what together with higher masses of BHs, give larger orbital speeds \citep[see also][]{Renzo1904}.

A distribution of \ivbh\ for standard NK prescription \citep[e.g. \std\ model;][]{Hobbs0507} has two distinctive parts (Fig.~\ref{fig:dSBH_kick}). The low-velocity peak is connected with BHs formed with no NK due to a heavy fallback, which, as we assume, decreases the NK, or in a direct collapse associated with no NK. The wide high-velocity component comprises BHs which are much lighter, thus formed with no significant fallback. Models \nkr\ and \nkbe\ lead to a comparable number of dSBHs ($6.4\times10^7$ and $6.5\times10^7$) as the \std\ model ($5.4\times10^7$), however, the shape of \ivbh\ distribution is significantly different. \nkr\ model lacks a low-velocity peak as the NK is independent of the fallback. Therefore, the average NK is much higher than in the \std\ model and typically equals $\sim125$ km/s for a typical BH mass of $7.5\msun$. For \std\ model there is typically full fallback associated with the formation of BHs with masses of $\sim7$--$8\msun$ and no NK is applied. Even more dissimilar is the \ivbh\ distribution for the \nkbe\ model, where NKs are proportional to the ratio of ejected mass to compact object mass ($v_{\rm NK}\sim M_{\rm ej}/M_{\rm BH}$; eq.~\ref{eq:NK_BE}) and typically, for $7.5\msun$ BHs, are $\sim130$ km/s.

\subsection{Bound systems}\label{sec:bhb}

The number of BHs which are bound in binaries is between $\sci{5.7}{4}$ -- $\sci{2.1}{7}$ (depending on the model, see Tab.~\ref{tab:results}) per MWEG. The fraction of BHs which reside in binaries for most of the tested models is $\lesssim17\%$. For models \nkr\ and \nkbe, the fraction of BHs in binaries is $\lesssim0.1\%$, therefore, the model of NKs highly influences the survival of BH binaries. 

The maximum mass for a BH is $\sim22\msun$ (or $\sim45$ for lower metallicity models). The binary with such a massive BH exists for $\sim100\kyr$ before it disrupts after the second SN. Afterwards, these BHs become the most heavy dSBHs (Sec.~\ref{sec:SBH_mass}).

Most of the binaries harbouring BHs are wide, i.e. formed without any interactions between stars. BHs reside  predominantly in DCOs (Sec.~\ref{sec:dco}). The rest is mostly accompanied by WDs (such binaries we treat as non-DCOs, i.e. BHBs; Sec.~\ref{sec:bhb_wide}). A small fraction  of BHBs could be observed during their MT phase, although this fraction is strongly dependent on the adopted SFH (Sec.~\ref{sec:mt}).

\subsubsection{Double compact objects} \label{sec:dco}

\begin{deluxetable*}{l|ccrcrcrcrcrcr}
    \tablewidth{\textwidth}
    \tablecaption{Double compact objects containing black holes}
    \tablehead{ Model & $N_\mathrm{BH,DCO}$ & \multicolumn{2}{c}{BH+BH} & \multicolumn{2}{c}{\rdcoa} & \multicolumn{2}{c}{\rdcob} & \multicolumn{2}{c}{$a[\rsun]<10^3\rsun$} & \multicolumn{2}{c}{$t_\mathrm{merge}<10\gyr$}}
    \startdata
    \tiny
    \std & \sci{5.1}{6} & \sci{4.9}{6} & (95.9\%) & \sci{2.4}{6} & (90.1\%) & \sci{1.6}{4} & (0.6\%) & \sci{2.5}{4} & (0.9\%)  & \sci{1.7}{4} & (0.6\%) \\
    \midz & \sci{1.2}{7} & \sci{1.2}{7} & (96.6\%) & \sci{2.6}{6} & (42.6\%) & \sci{4.0}{5} & (3.4\%) & \sci{1.7}{6} & (27.3\%)  & \sci{8.1}{5} & (13.0\%) \\
    \lowz & \sci{1.9}{7} & \sci{1.9}{7} & (97.5\%) & \sci{5.7}{6} & (57.8\%) & \sci{1.8}{6} & (18.1\%) & \sci{2.7}{6} & (27.0\%)  & \sci{1.9}{6} & (19.0\%) \\
    \ssa & \sci{6.9}{6} & \sci{6.6}{6} & (95.6\%) & \sci{3.4}{6} & (93.1\%) & \sci{2.4}{4} & (0.7\%) & \sci{3.0}{4} & (0.8\%)  & \sci{2.3}{4} & (0.6\%) \\
    \nkr & \sci{1.1}{4} & \sci{5.1}{3} & (45.5\%) & \sci{8.8}{2} & (10.3\%) & \sci{1.6}{3} & (19.1\%) & \sci{6.8}{3} & (79.4\%)  & \sci{3.2}{3} & (36.8\%) \\
    \nkbe & \sci{2.7}{3} & \sci{2.5}{2} & (9.5\%) & \multicolumn{2}{c}{--} & \multicolumn{2}{c}{--} & \sci{2.5}{3} & (100\%)  & \sci{1.3}{2} & (5.0\%) \\
    \imff & \sci{1.3}{7} & \sci{1.3}{7} & (96.6\%) & \sci{6.2}{6} & (90.6\%) & \sci{3.8}{4} & (0.6\%) & \sci{6.0}{4} & (0.9\%)  & \sci{3.8}{4} & (0.6\%) \\
    \imfs & \sci{1.6}{6} & \sci{1.5}{6} & (95.3\%) & \sci{7.7}{5} & (90.7\%) & \sci{3.7}{3} & (0.4\%) & \sci{5.7}{3} & (0.7\%)  & \sci{3.7}{3} & (0.4\%)
    \enddata
    \tablecomments{Results for double compact objects harbouring BHs (i.e. BH+BH and BH+NS systems). $N_\mathrm{BH,DCO}$ - number of BHs in double compact objects; BH+BH - number of BHs in double BH systems (the rest reside in BH+NS systems); \rdcoa\ and \rdcob\ - number of BHs formed through specific evolutionary routes; $a[\rsun]<10^3\rsun$ - number of BHs residing in close DCOs, i.e. systems with separations lower than $10^3\rsun$; $t_\mathrm{merge}<10\gyr$ - number of BHs in systems with time to merger after second supernova smaller than $10\gyr$.}
    \label{tab:dco}
\end{deluxetable*}

In this study, by DCOs we understand only these harbouring at least one BH (i.e. BH+BH and BH+NS), totally neglecting NS+NS systems. BH+WD systems are included in non-DCO binaries (BHBs; see Sec.~\ref{sec:bhb_wide}). The fact that the majority of binaries harbouring BHs are actually DCOs is a direct consequence of the adopted uniform distribution of mass ratios on ZAMS - if the primary is a BH progenitor ($\mzams\gtrsim20\msun$) the most likely companions are NS and BH progenitors ($\mzams\gtrsim8\msun$). NSs generally receive higher NKs than BHs and many potential BH+NS binaries get disrupted during the NS formation \citep[e.g.][]{Dominik1211}, therefore, BH+BH systems typically dominate ($\sim95$--$98\%$; see Tab.~\ref{tab:dco}) over BH+NS binaries in most tested models. The exception are models with higher average NKs (\nkr\ and \nkbe) where most of DCOs are compact BH+NS systems. Many groups studied the formation of DCOs harbouring BHs over the years \citep[e.g.,][]{Lipunov9706,Nelemans0109,Belczynski0206,Voss0307,Dominik1211,Belczynski1606,Belczynski1610,Eldridge1611,Stevenson1704}. However, the main focus of those studies was lied on the merger rates and properties of gravitational wave sources and, hence, they performed no deeper analysis of wider DCOs. Here, we present the analysis of the entire DCO population as a part of the total BH population. Therefore, our results may be applied not only to the study of merging DCOs, but also to other phenomena, e.g. microlensing by DCOs. 

In Tab.~\ref{tab:dco} we present detailed results for the main models. The number of BHs in DCOs is typically between \sci{1.5}{6} -- \sci{1.9}{7} (note that BH+BH counts as two BHs), although for \nkr\ and \nkbe\ models, the number drops to \sci{5.1}{3} and \sci{2.5}{2}, respectively. The low number of DCOs in the models with higher average NKs is connected to the fact that most of binaries with massive components which are potential progenitors of DCOs are wide (route \rdcoa; see below), thus easily disrupted by even moderate NKs.

\begin{figure}[t]
    \centering
    \includegraphics[width=1.0\columnwidth]{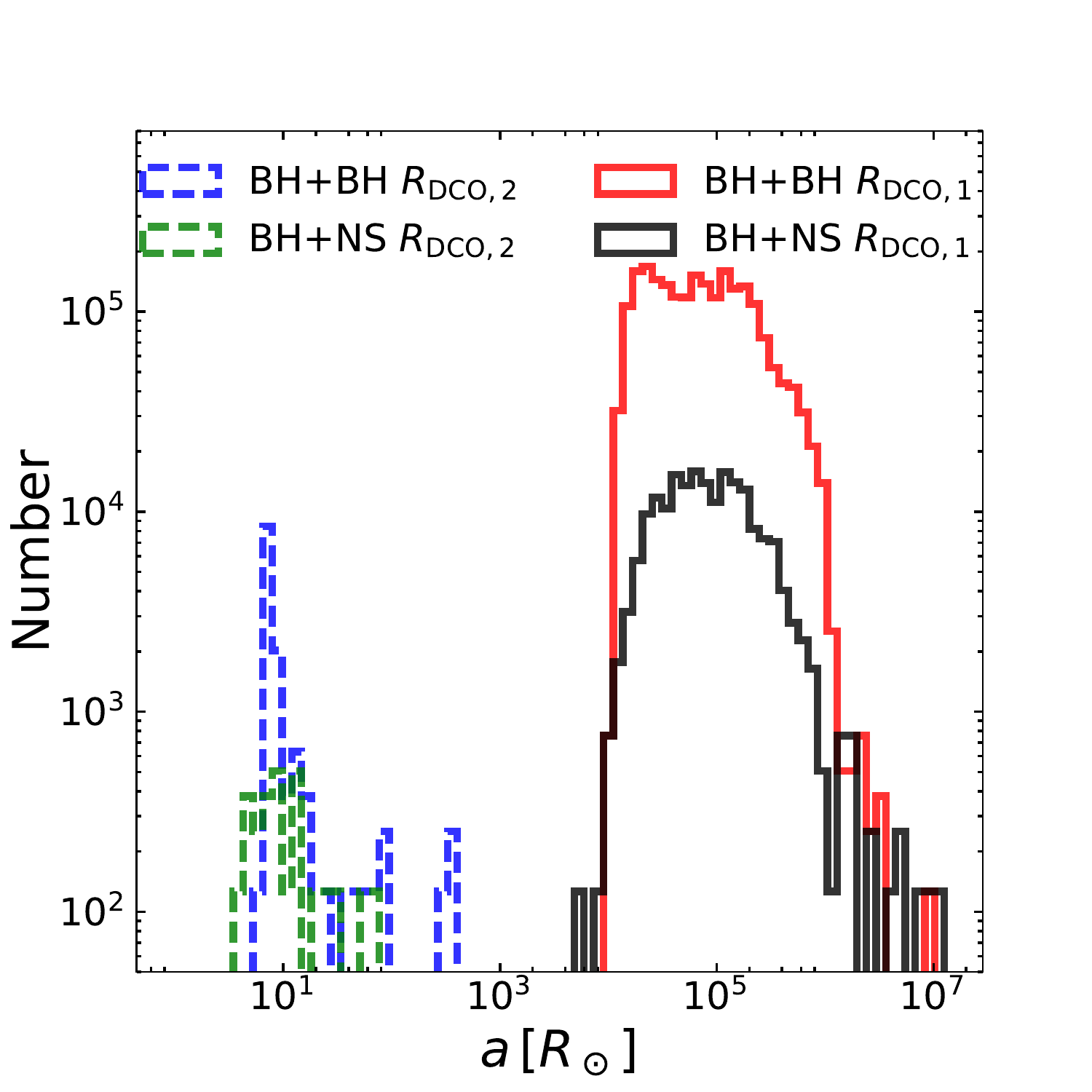}
    \caption{Distribution of separations of double compact objects at the moment of their formation for the \std\ model. Presented are distributions for BH+BH and BH+NS binaries and two main evolutionary routes \rdcoa\ and \rdcob. A gap at $\sim10^3\dash10^4\rsun$ separates double compact objects originating from different evolutionary channels. Large separations are a property of systems that were formed without any binary interactions (\rdcoa\ route). Double compact objects formed in \rdcob\ route have smaller separation ($a\lesssim10^3\rsun$), what is a direct consequence of a CE phase during the formation process.}
    \label{fig:dco_a}
\end{figure}

There are two main evolutionary routes leading to the formation of DCOs harbouring BHs (Tab.~\ref{tab:er}). The main feature that distinguish these routes is the initial separation. It is connected with the lack (\rdcoa), or presence (\rdcob) of the pre-DCO formation interactions (MT and CE), which results in the formation of wide ($a\gtrsim10^4$), or close ($a\lesssim10^3\rsun$) DCOs, respectively (see Fig.~\ref{fig:dco_a}). Among close DCOs (mainly route \rdcob) are potential DCO merger progenitors \citep[e.g.][]{Abbott1602} and their number and relative fraction strongly depends on metallicity and NKs as previously noted by \citet{Chruslinska19}.

\begin{figure}[t]
    \centering
    {\large\rdcoa}\\
    \includegraphics[width=1.0\columnwidth]{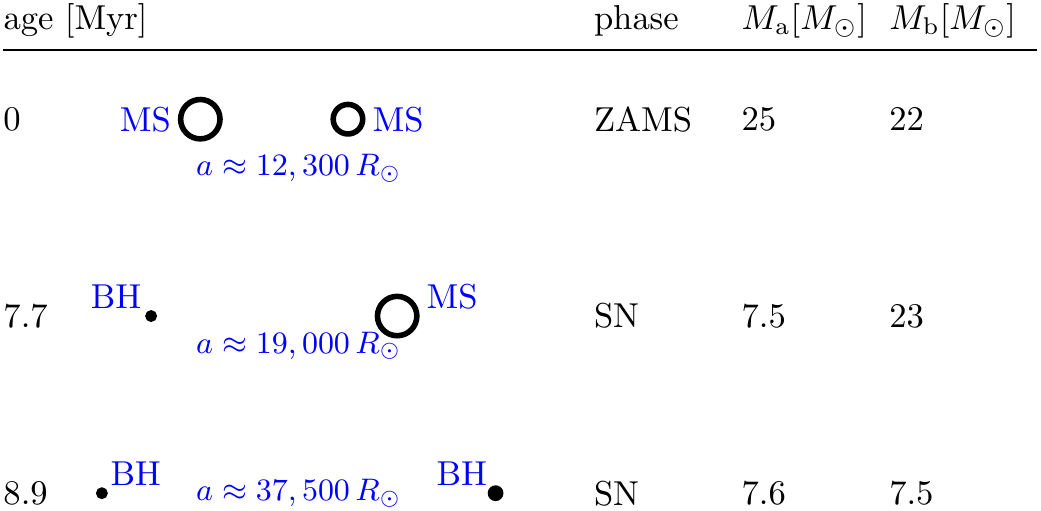}\\
    \vspace{0.5cm}
    {\large\rdcob}\\
    \includegraphics[width=1.0\columnwidth]{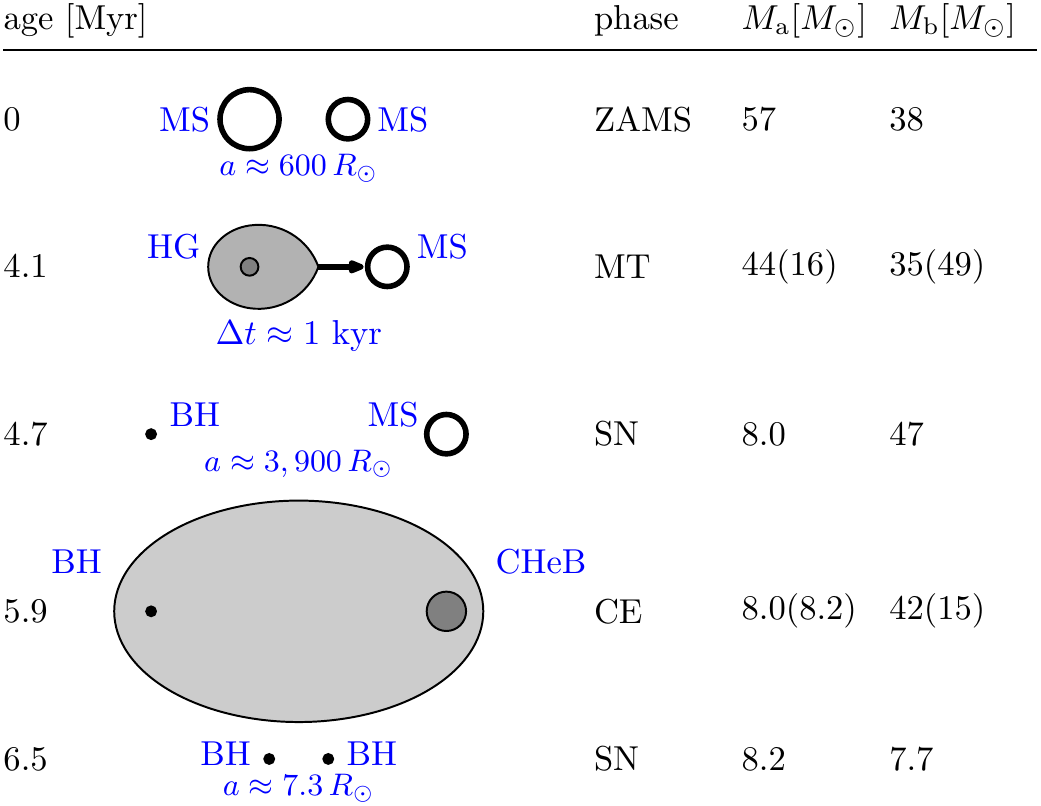}
    \caption{Main evolutionary phases leading to the formation of DCOs harboring BHs. Route \rdcoa\ is presented on the upper plot, whereas route \rdcob\ on the lower one. For explanations of the majority of abbreviations see Fig.~\ref{fig:rdSBH}. Additionally: CE - common envelope; CHeB - core helium burning.}
    \label{fig:rdco}
\end{figure}

Route \rdcoa\ is the main formation route of DCOs if standard NK model is concerned. In this formation channel none of strong interactions are present (Fig.~\ref{fig:rdco}). In a typical situations the stars on ZAMS are not very massive ($\sim25$ and $\sim22\msun$) and the separation is large ($\sim12,300\rsun$). After about $7.7\myr$ the primary explodes and forms a $\sim7.5\msun$ BH. In additional $\sim1\myr$, the secondary forms a $\sim7.5\myr$ BH in a SN explosion. The final separation is $\sim37,500\rsun$.

The other formation route (\rdcob) is responsible for much smaller fraction of DCOs ($\lesssim21\%$), but systems which form through this route have much shorter separations and, thus, may merger after less than $10\gyr$ due to the emission of gravitational waves. In a typical case, the binary is massive on ZAMS ($\mzamsa\approx57\msun$ and $\mzamsb\approx38\msun$) and the separation is small ($\azams\approx600\rsun$). The primary evolves fast and after $\sim4.1\myr$ fills its Roche lobe while expanding on the HG. The MT commences and results in the loss of nearly entire hydrogen envelope by the primary, which shortly after becomes a $\sim16\msun$ HeS. Half of the primaries envelope is accreted by the secondary which grows to $\sim49\msun$ while still on its MS. After additional $\sim600\kyr$, the primary forms a $\sim8\msun$ BH in a SN explosion. The secondary starts to expand after leaving the MS and while being a CHeB star fills its Roche lobe and a CE commences. As a result the separation shrinks from $\sim4000\rsun$ to about $5.2\rsun$ and the secondary loses its entire envelope and becomes a $\sim15\msun$ HeS. A second SN occurs $\sim600\kyr$ later and $\sim7.7\msun$ BH forms. The NK is not very strong, thus the separation remains small ($\sim7.3\rsun$). Such a binary is estimated to merge after $\sim420\myr$. This channel was previously thoroughly analysed in the context of DCO mergers by e.g. \citet{Belczynski0206,Dominik1211,Belczynski1606,Woosley1606,Kruckow1812}.

BH+NS formation through \rdcoa\ and \rdcob\ routes is qualitatively similar to the formation of BH+BH systems, although the ZAMS secondary masses are significantly lower ($\sim3\times$ for \rdcoa\ and $\sim1.5\times$ for \rdcob). Additionally, the progenitors are more easily disrupted during the second SN and have a smaller chance of CE survival due to more extreme mass ratio. This results in a significantly lower fraction of BH+NS systems among DCOs than BH+BH systems, in spite of a higher abundance of potential progenitors in initial (ZAMS) populations\footnote{The ratio of potential BH+NS progenitors to potential BH+BH progenitors (i.e. binaries with components' masses high enough to form a NS, or BH in single star evolution) may be calculated as
\[
f=\frac{\int_{M_{\rm NS,ZAMS,max}}^{150\msun}P_{\rm BH+NS}(M_1)\,{\rm IMF}(M_1)\,dM_1}{\int_{M_{\rm NS,ZAMS,max}}^{150\msun}P_{\rm BH+BH}(M_1)\,{\rm IMF}(M_1)\,dM_1},
\]
where $P_{\rm BH+NS/BH+BH}(M_1)$ is the probability that a companion of a BH progenitor with mass $M_1$ is a NS/BH progenitor. Assuming flat mass ratio distribution for companions ($P(q)=1/(M_1-M_{\rm NS,ZAMS,min})$),
\[
P_{\rm BH+NS}(M_1)=\frac{M_{\rm NS,ZAMS,max}-M_{\rm NS,ZAMS,min}}{M_1-M_{\rm NS,ZAMS,min}}\] and \[P_{\rm BH+BH}(M_1)=\frac{M_1-M_{\rm NS,ZAMS,max}}{M_1-M_{\rm NS,ZAMS,min}},
\]
where $M_{\rm NS,ZAMS,min/max}=8/22\msun$ are the minimal/maximal ZAMS masses for a NS progenitor (assuming no interactions with the companion) for solar metallicity (\zsun). ${\rm IMF}(M_1)\propto M_1^\Gamma$ is the initial mass function. $150\msun$ is the upper limit for the ZAMS mass adopted in our calculations. Assuming $\Gamma=-2.3$ and solar metallicity (\std\ model) $f\approx1.1$ meaning more BH+NS than BH+BH progenitors in initial populations. This value is slightly affected by metallicity, as $Z=1\%\zsun$ (\lowz\ model; where $M_{NS,ZAMS,max}\approx19\msun$) gives $f\approx1$, and significantly affect by the steepness of the IMF, as $\Gamma=-1.9$ (\imff\ model) gives $f\approx0.9$, i.e. more BH+BH progenitors.}. There are typically (except \nkr\ and \nkbe\ models) $\gtrsim20$ times more BHs in BH+BH binaries than in BH+NS. We observe no significant difference between the shapes of BH+BH and BH+NS distributions of orbital separations at the moment of of DCO formation (Fig.~\ref{fig:dco_a}).

We have found that the number of BHs in DCOs is $3$--$4$ times higher in lower metallicity environments than in the \std\ model. The number of wide DCOs (\rdcoa) is only slightly affected by metallicity (by a factor of $\lesssim2.5$), whereas the number of close DCOs (\rdcob) is significantly affected (by more than two orders of magnitude). The latter is an effect of higher BH masses (typically $\sim15$) in low metallicity environments in comparison to \zsun\ environments (typically $\sim7$--$8\msun$). Heavier BHs in the standard NK model usually obtain lower NKs due to fallback, or no NK when formed in a direct collapse. Therefore, DCOs in lower metallicity environments, where BHs form with higher masses, are less frequently disrupted \citep[e.g.][]{Belczynski1006}.

Significantly different situation concerns the relation for close systems (\rdcob), which in lower metallicity environments are about two orders of magnitude more numerous than in the \std\ model. This results from an interplay between mass loss in stellar wind and stellar expansion (Fig.~\ref{fig:exp-Z}). The former is proportional to metallicity and leads to orbital widening due to mass loss from the system. On the other hand, solar metallicity stars do not expand as significantly as these in \midz\ model. This means that massive binaries in high-metallicity environment (e.g. \std\ model) more frequently have separations too large to go through a CE phase which is an essential step in the \rdcob\ route. Consequently, lower-metalicity models have more close DCOs, than high metallicity ones. A similar result was recently obtained by \citet{Spera1905} with the use of SEVN population synthesis code.

\begin{figure}[t]
    \centering
    \includegraphics[width=1.0\columnwidth]{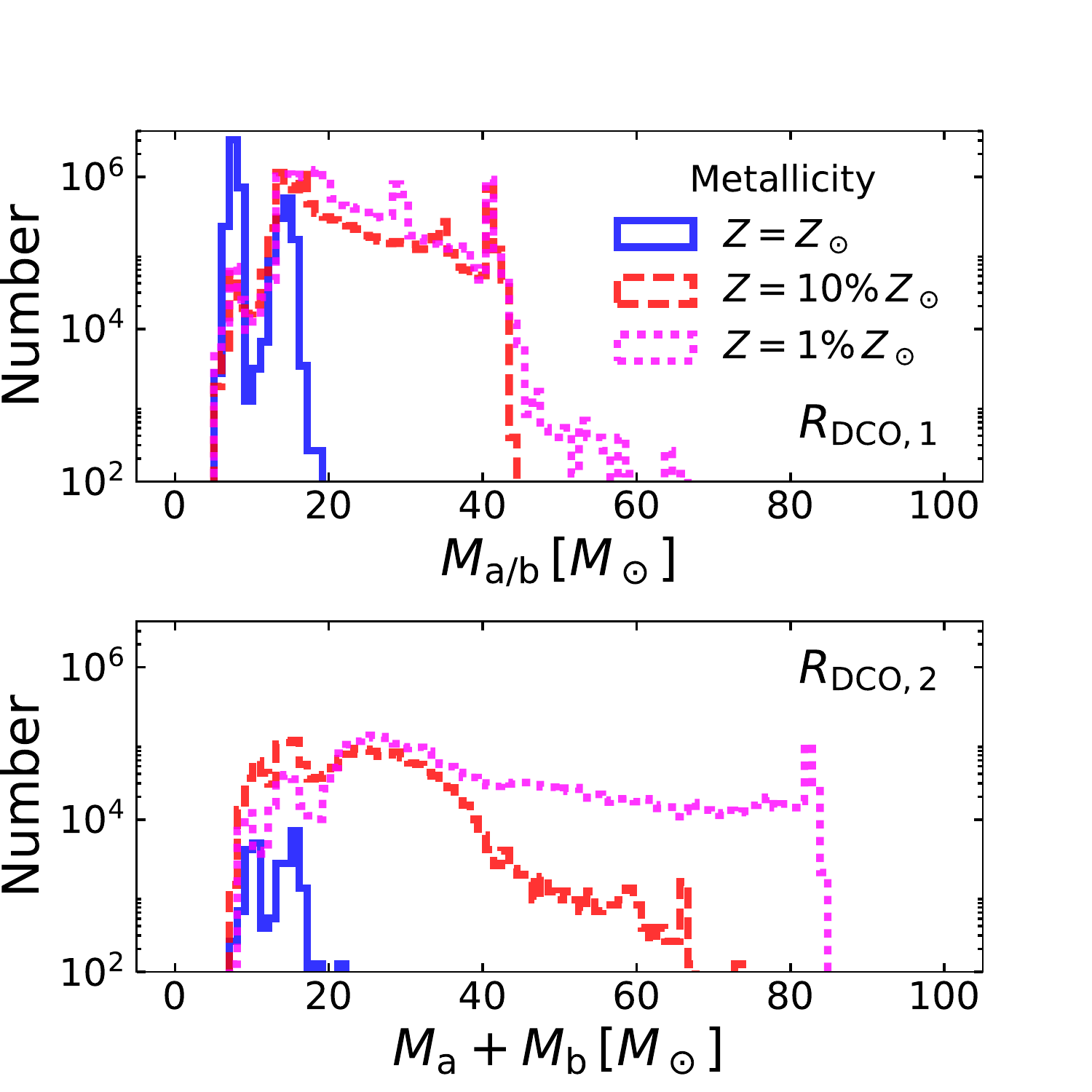}
    \caption{Distribution of components' masses of wide double compact objects ($a>10^3\rsun$; route \rdcoa; upper plot) and total binary masses for close double compact objects ($a\leq10^3\rsun$; route \rdcob; lower plot) scaled for MWEG. Different metallicites are presented: $Z=\zsun$ (\std\ model), $Z=10\%\,\zsun$ (\midz\ model), and $Z=1\%\,\zsun$ (\lowz\ model). In case of wide BH+NS systems (upper plot), only BH masses are presented.}
    \label{fig:dco_mass}
\end{figure}

Distributions of component masses are shown in Fig.~\ref{fig:dco_mass}. For \rdcoa\ formation channel (wide DCOs) distributions for primaries and secondaries are joined together, except BH+NS systems for which only BH masses are shown. The results resemble these for BHs from disrupted binaries (Fig.~\ref{fig:dSBH_mass}). For \rdcob\ channel (close DCOs) the total mass of a binary is provided. Such information is more important from the point of view of microlensing surveys, because a close binary will most probably act as a single lens. For lower metallicites the total mass may reach $\sim70$--$80\msun$ what matches the current range of observed mass in BH+BH mergers \citep[see also \citeauthor{Belczynski1606} \citeyear{Belczynski1606}]{LIGO1811}.

Within our models up to $\sim15\%$ of BHs are found in DCOs (Tab.~\ref{tab:results}) which are predominantly wide \citep{Voss0307}. For the standard NK model, up to $19\%$ of them have merger times ($t_{\rm merge}$) smaller than $10\gyr$, what mainly depends on metallicity \citep[see Tab.~\ref{tab:dco}; merging BH+BH systems form much more efficiently at low Z, e.g.][]{Dominik1312,Giacobbo1810,Klencki1811,Chruslinska19}. In the \nkr\ and \nkbe\ models, although the fraction of merging systems may be much higher ($\sim37\%$ for \nkr\ model), the number of these systems is significantly smaller. It is a consequence of higher average NKs ($\sim125$--$130$ km/s in contrast to $\sim10$ km/s in \std\ model), what frequently disrupts wide binaries. 

\begin{figure}[t]
    \centering
    \includegraphics[width=1.0\columnwidth]{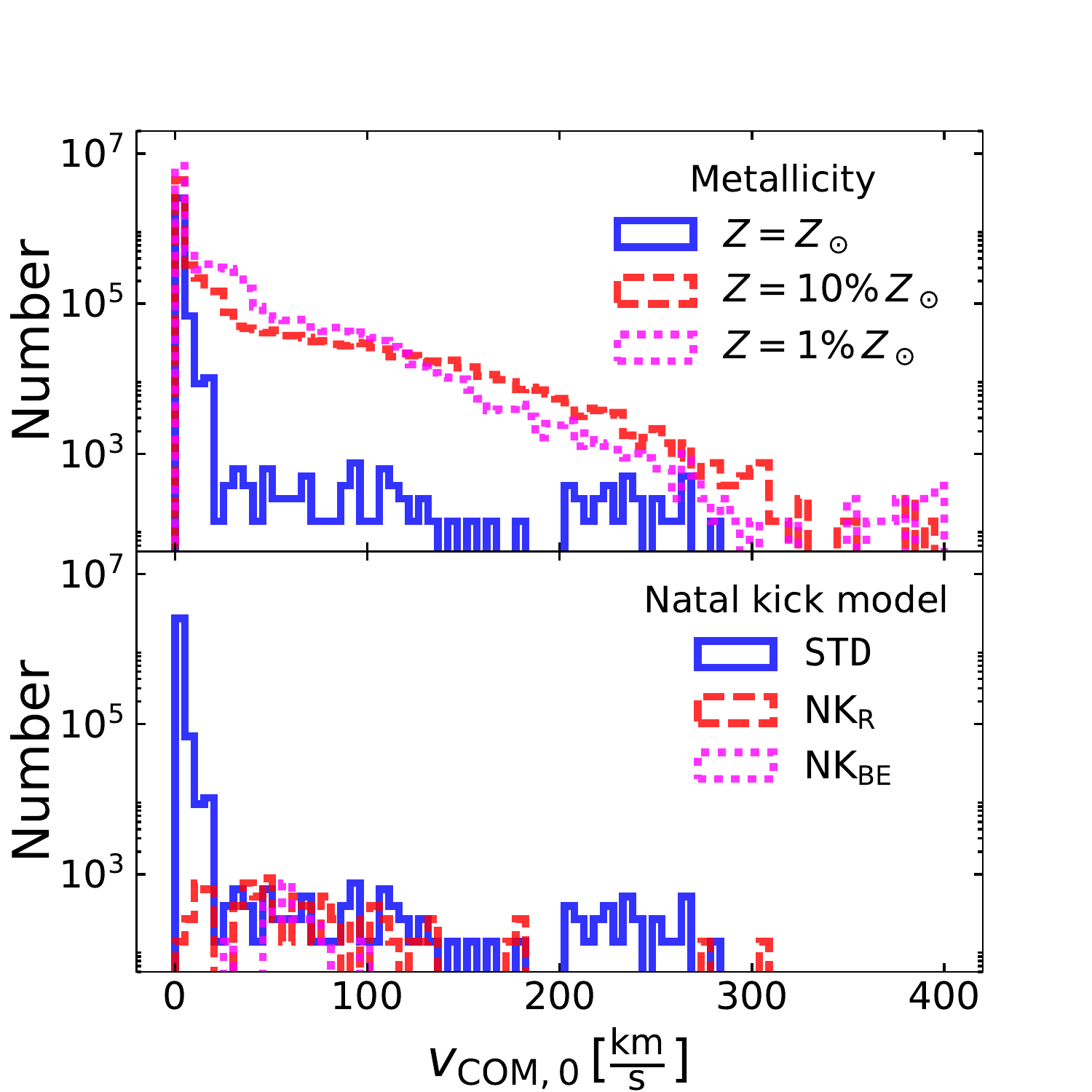}
    \caption{Distribution of the initial center-of-mass velocities of double compact objects just after the formation of the second compact object ($v_{\rm COM,0}$). Note that the motion in the gravitational potential is not included here. Upper plot compares distributions for different metallicities, whereas the lower one compares different NK models.}
    \label{fig:dco_v}
\end{figure}

Typical center-of-mass initial velocities ($v_{\rm COM,0}$; i.e. binary velocities just after the second SN not including velocity resulting from the motion in a gravitational potential) are below $\sim20$ km/s (Fig.~\ref{fig:dco_v}). Only less than $15\%$ of DCOs (mainly compact; route \rdcob) possess velocities exceeding $20$ km/s. Therefore, vast majority of the DCOs remains in the vicinity of their birth places. Highest velocities ($v_{\rm COM,0}\gtrsim200$ km/s) are obtain by BH+NS systems, whereas systems with lower velocities are dominated by BH+BH binaries, what results from typically higher NKs obtained by NSs, than BHs. Typical velocities of merging systems ($t_{\rm merge}<10\gyr$) are below $\sim50$ km/s for \std\ model, so most of the mergers are predicted to occur in the vicinity of the formation places \citep[e.g.][]{Perna1807}. However, some systems may obtain also high velocities ($\gtrsim200$ km/s). These are mainly BH+NS systems, which we predict to merge away from their birth environments. For \nkr\ and \nkbe\ models, DCOs obtain typically intermediate velocities ($\sim50$--$130$ km/s; mainly route \rdcob; typically, $t_{\rm merge}\lesssim10\gyr$), although some systems (mainly BH+NS) may have velocities as high as $\sim400$ km/s.

\subsubsection{Wide non-DCO binaries}\label{sec:bhb_wide}

\begin{deluxetable*}{l|ccc|ccc}
    \tablewidth{\textwidth}
    \tablecaption{Wide and mass-transferring binaries}
    \tablehead{ Model & \multicolumn{3}{c}{Wide BHB} & \multicolumn{3}{c}{MTBHB} \\ & $N$ & $a[\rsun]$ & Typ. comp. &  $N$ & $a[\rsun]$ & Typ. comp.}
    \startdata
    \std  & \sci{1.5}{6}  & $\gtrsim10,000$ & CO WD ($86\%$) & \sci{8.5}{1} & $\lesssim115$  & MS ($75\%$) \\
    \midz & \sci{7.5}{5}  & $\gtrsim8,300$  & CO WD ($79\%$) & \sci{4.2}{3} & $\lesssim110$ & MS ($85\%$) \\
    \lowz & \sci{1.2}{6}  & $\gtrsim5,800$  & CO WD ($85\%$) & \sci{4.4}{2} & $\lesssim84$  & MS ($91\%$) \\
    \ssa  & \sci{3.3}{6}  & $\gtrsim8,300$  & CO WD ($74\%$) & \sci{5.7}{2} & $\lesssim108$  & MS ($73\%$) \\
    \nkr  & \sci{3.1}{3}  & $\gtrsim4,800$  & CO WD ($89\%$) & \sci{1.7}{2} & $\lesssim90$  & MS ($86\%$) \\
    \nkbe & \sci{1.0}{-4} & $\gtrsim14,000$ & CHeB ($100\%$)  & \sci{1.8}{2} & $\lesssim75$  & MS ($82\%$) \\
    \imff & \sci{3.1}{6}  & $\gtrsim10,000$ & CO WD ($86\%$) & \sci{1.3}{2} & $\lesssim113$  & MS ($73\%$) \\
    \imfs & \sci{5.5}{5}  & $\gtrsim10,000$ & CO WD ($86\%$) & \sci{3.2}{1} & $\lesssim116$  & MS ($70\%$) \\
    \enddata
    \tablecomments{Results for wide and mass transferring binaries with one of the components being a BH. $N$ - number of binaries (also number of BHs); $a$ - typical separation; Typ. comp. - typical evolutionary type of the companion: MS - main sequence; CHeB - core helium burning; CO WD - carbon-oxygen white dwarf.}
    \label{tab:bhb}
\end{deluxetable*}

Throughout the paper we use the term wide BHB referring to systems harbouring a BH with a non-compact companion and experiencing no strong interaction during their entire evolution (the separation between the stars is large and the stars never fill their Roche lobes). In this section, we focus on wide BHBs, excluding wide DCOs, which were discussed in previous section. 

Tab.~\ref{tab:bhb} summarizes our results for the main models. The typical number of wide BHBs is between $\sci{5.5}{5}$ and $\sci{3.3}{6}\msun$. Only for models with higher average NKs (\nkr\ and \nkbe) the predicted number of such binaries is much lower (\sci{3.1}{3} and \sci{1.0}{-4}, respectively). The probability for a binary to obtain a low NK ($\lesssim20$ km/s) in \nkr\ and \nkbe\ models is much lower than in \std\ model (c.f. Fig.~\ref{fig:dSBH_kick}) and typically NKs are $\gtrsim100$ km/s. Therefore, in these models wide binaries are more easily disrupted during the formation of a BH. In contrast, standard NK model \citep{Hobbs0507} predicts mostly low, or negligible NKs for BHs which allows for more frequent survival of wide binaries.

The number of wide BHBs is higher in \std\ model than in \midz\ and \lowz\ models mainly because of the more significant orbital expansion resulting from mass loss in stellar wind, which is stronger for higher metallicities. Additionally, stellar expansion is higher in \midz\ model, thus futher increasing the chance of interactions (Fig.~\ref{fig:exp-Z}). In \lowz\ model the stellar expansion is smaller than in the \midz\ model, therefore, results in higher number of wide BHBs.

BH companions in wide systems are predominantly CO WDs ($74$--$89\%$). WDs, being the final evolutionary stage of low-mass ($\lesssim8\msun$) stars' evolution, are simultaneously the longest evolutionary stage of the evolution of stars with initial masses $2\lesssim M\lesssim8\msun$ during the Hubble time. For comparison, low mass stars ($M\lesssim1\msun$) spend most of this time on the MS, whereas, heavier stars ($M\gtrsim8\msun$) quickly ($\lesssim50\myr$) end their evolution and form a NS, or a BH. The typical separations of wide BHBs are $a\gtrsim5000$.

\begin{figure}[t]
    \centering
    \includegraphics[width=1.0\columnwidth]{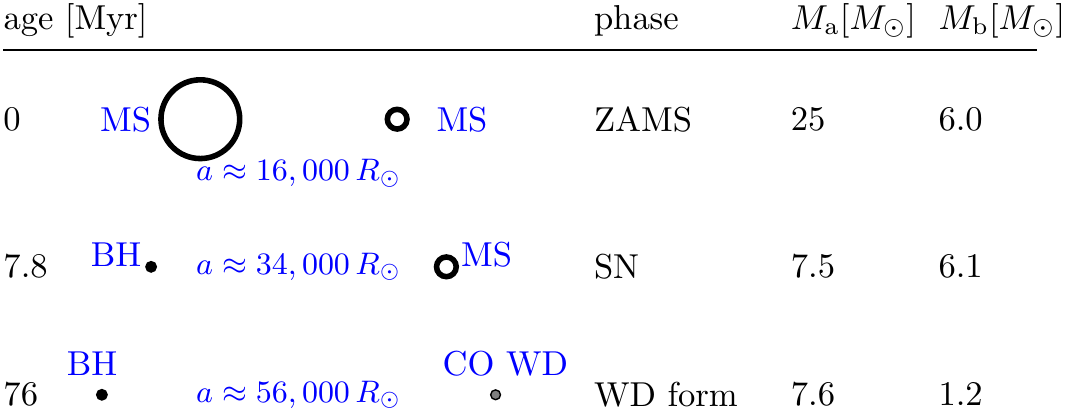}
    \caption{Evolution towards the formation of a typical wide binary containing a BH (see Sec.~\ref{sec:bhb_wide}). For descriptions of typical abbreviations and parameters see Fig.~\ref{fig:rdSBH}. Additionally: WD form - formation of a white dwarf; CO WD - carbon-oxygene white dwarf.}
    \label{fig:rwide}
\end{figure}

In an exemplary evolution of a typical wide BHB, the progenitor system on ZAMS is composed of a $\sim25\msun$ primary and a $6.0\msun$ companion (see Fig.~\ref{fig:rwide}). 
The initial separation is large ($\sim16,000\rsun$), which precludes any interactions. 
The separation further increases due to wind mass-loss from the stars and just before the SN reaches $\sim34,000\rsun$. The primary, being heavier, evolves faster and in $\sim7.8\myr$ becomes a BH. The NK and the Blaauw kick are crucial at that point. Only a small total kick will allow for a survival of the binary. The secondary is still on its MS for another $\sim60\myr$.  At the age of about $76\myr$ a CO WD forms. The final separation may reach $56,000\rsun$.

\begin{figure}[t]
    \centering
    \includegraphics[width=1.0\columnwidth]{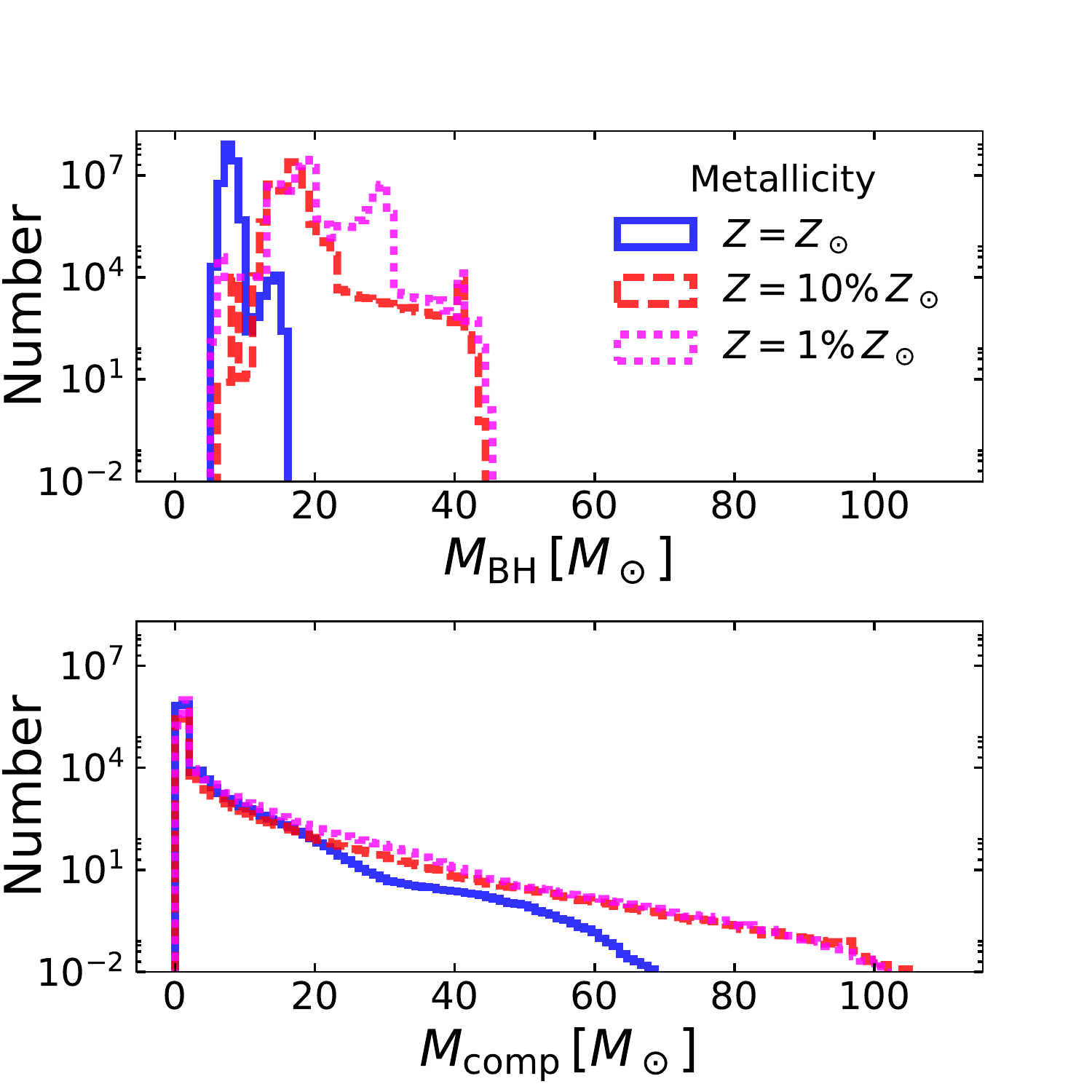}
    \caption{Distribution of BH ($M_{\rm BH}$; upper plot) and companion ($M_{\rm comp}$; lower plot) masses for wide BHBs. Results for three metallicites are presented: $Z=\zsun$ (\std\ model), $Z=10\%\,\zsun$ (\midz\ model), and $Z=1\%\,\zsun$ (\lowz\ model). The peak in the companions mass distribution is composed of WDs, whereas the tail is mostly formed of MS stars.}
    \label{fig:wide_mass}
\end{figure}

Mass distribution of BHs in wide BHBs (Fig.~\ref{fig:wide_mass}) resembles this for dSBHs as most dSBHs are not interacting significantly prior to disruption, thus frequently originate from wide BHBs. Two distributions differ in the presence (dSBHs), or the lack (wide BHBs) of the high mass peak ($\sim22\msun$) for \std\ model. Also the high-mass tail (above $\gtrsim15\msun$ for \std\ model and $\gtrsim40\msun$ for \midz\ and \lowz\ models), which originates from wind accretion by a BH, or its progenitor, in the case of non-interacting binaries is less pronounced, than in distributions for interacting binaries. The maximal BH masses are $\sim16\msun$ for \std\ model and $\sim46\msun$ in lower metallicities. The distribution of companion masses is dominated by WDs (carbon-oxygen and oxygen-neon), which form a characteristic peak between $\sim1\msun$ and the Chandrasekhar mass ($M_{\rm Ch}=1.44\msun$). The long tail on this distribution is mainly composed of MS stars, which may have masses up to $\sim80\msun$ for solar metallicity, or $\gtrsim100\msun$ for lower metallicity models. The massive companions ($M\gtrsim8\msun$) are going to explode in SN explosion which may disrupt the BHBs (either through a NK, or a Blaauw kick) populating dSBHs, or form a wide DCOs (discussed earlier in Sec.~\ref{sec:dco}).

Although in rare cases the initial center-of-mass velocity ($v_{\rm COM,0}$) of wide BHBs may reach $\sim80$ km/s, typically  ($\gtrsim95\%$ of cases) they are negligible ($\lesssim10$ km/s). Due to the fact that orbital velocities are typically also small ($\lesssim 20$ km/s), the binaries rarely leave their birth places if evolved in isolation. We note, that in realistic situation a wide BHB is frequently influenced by fly-by encounters \citep[e.g.][]{Klencki1708}.

\subsubsection{Mass Transferring binaries with BH accretors}\label{sec:mt}

Mass transferring BHBs (MTBHB) have different evolutionary routes and properties than the typical BHBs, which are dominated by non-interacting systems (wide BHBs; Sec.~\ref{sec:bhb_wide}). Although we note that BHs were detected also in high-mass XRBs where MT rates high enough to fuel an accretion disk may occur also through companion's stellar wind \citep[e.g.][]{Ruhlen1112}, in this study only RLOF systems were included.

For any adopted SFH the fraction of BHs accreting mass from their companions through RLOF is small. However, this small fraction is more a result of the brevity of MT phase in comparison to the evolutionary time-scales, than the rarity of BHBs that are close enough to commence RLOF. Here, for the sake of presentation, we assumed a constant star formation for the last $10\gyr$. In such a case, only $<1\%$ (see Tab.~\ref{tab:bhb}) of all BHBs contain a donor which is filling its Roche lobe and transferring mass onto a BH. MTBHBs can be perceptible as an X-ray binaries \citep[e.g.][]{Tauris06}. We found out that in our models $2$ -- $28\%$ of BHs have gone through a MT phase which lasts typically $\lesssim40\myr$. A BH accretes on average $0.7$ -- $3.1\msun$ in that time. 

Our simple model predicts $32$ -- $4,200$ MTBHBs per MWEG. The number is the highest in lower metallicities ($\sci{4.4}{2}$ and $\sci{4.2}{3}$ systems for \lowz\ and \midz\ models, respectively) where it is easier to start RLOF MT \citep{Linden1012}. Also in \ssa\ model we see $\sim8$ times higher predicted number of MTBHBs, than in the \std\ model. It is a result of assumed in this model thermal distribution of eccentricities ($P(e)\sim e$), what significantly increases average initial eccentricity, thus lowering the initial periastron distance, what makes RLOF easier.

There are two main evolutionary routes (Tab.~\ref{tab:er}) typical for all models (comprising $79\div99\%$ of all MTBHBs). Both channels differ mainly in the mass of an accreting BH. In the \rlmxba\ route, typical BH masses are above $\sim5\msun$, whereas in \rlmxbb\ route, they are around $\sim3\msun$. This distinctness stems from the different BH formation mechanisms. In \rlmxba, a BH forms directly after a SN, whereas in \rlmxbb, a NS forms first and after a period of mass accretion reaches a critical mass ($M_{\rm max,NS}$) and collapses to a BH. The critical mass is in general not well constrained and depends on the applied equation of state and rotation \citep[e.g][]{Kalogera9610}. In our simulations a value of $M_{\rm max,NS}=2.5\msun$ was adopted. 

\begin{figure}[t]
    \centering
    {\large\rlmxba}\\
    \includegraphics[width=1.0\columnwidth]{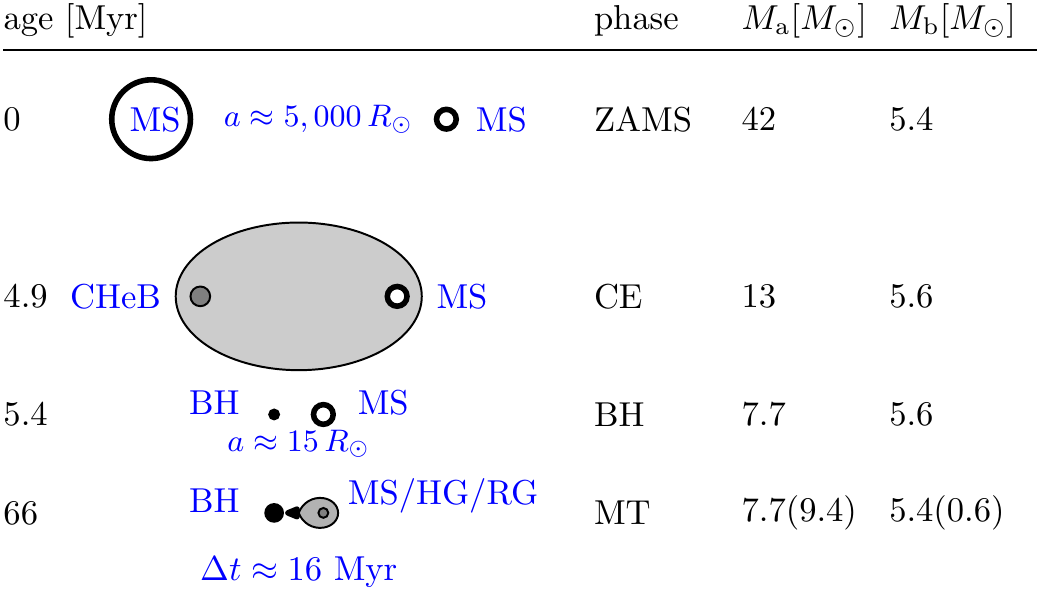}\\
    \vspace{0.5cm}
    {\large\rlmxbb}\\
    \includegraphics[width=1.0\columnwidth]{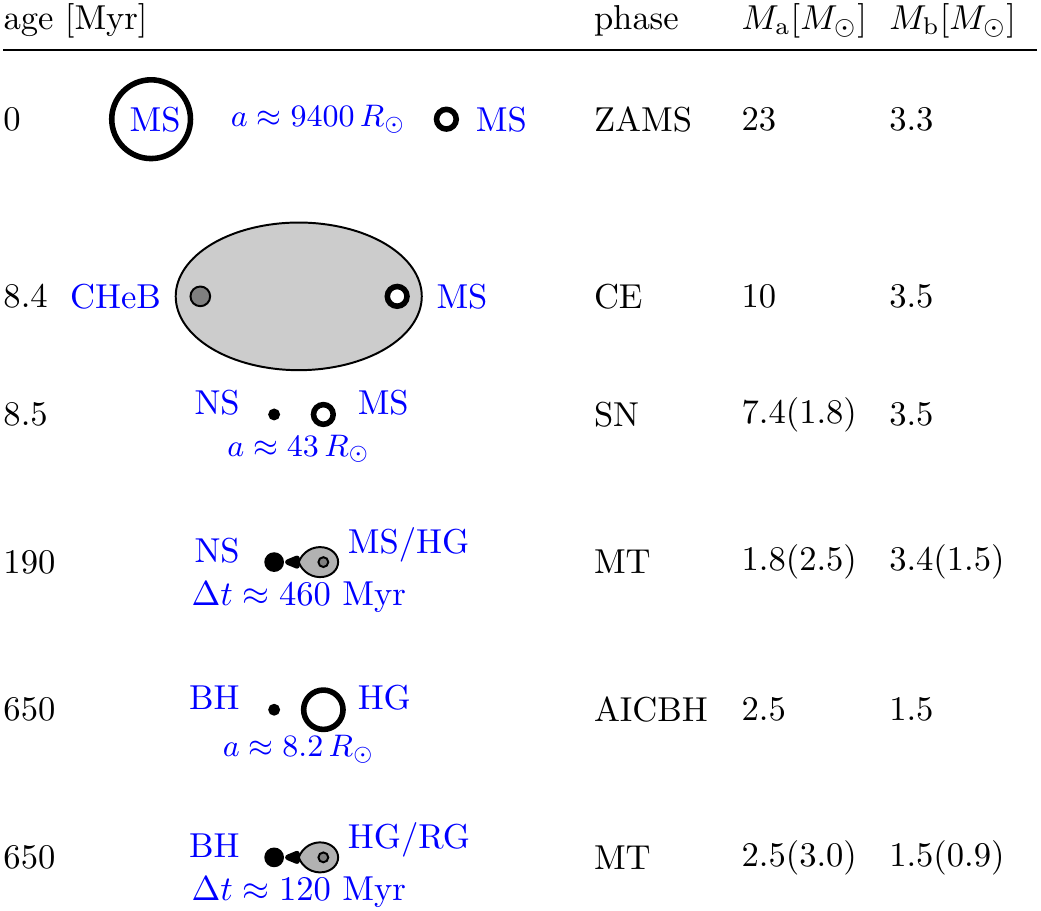}
    \caption{Evolution towards the formation of a typical MTBHB (route \rlmxba\ and \rlmxbb). For descriptions of abbreviations and parameters see Fig.~\ref{fig:rdSBH}. Additionally: CHeB - core helium burning; RG - red giant; CE - common envelope; AICBH - accretion induced collapse of a NS into a BH.}
    \label{fig:rlmxb}
\end{figure}

An exemplary system which becomes a MTBHB through \rlmxba\ route begins its evolution having ZAMS masses of $42\msun$ and $5.4\msun$ (Fig.~\ref{fig:rlmxb}, upper plot). The separation is moderate, $a_{\rm ZAMS}\approx5000\rsun$. In $\sim4.9\myr$, the primary evolves off the MS and commences a CE phase. Afterwards, the separation is reduced to $\sim13\rsun$. After additional $0.5\myr$, the primary becomes a BH receiving a small NK. The companion needs additional $\sim60\myr$ to expand due to nuclear evolution and fill its Roche lobe. The MT prolongs for $16\myr$ during which the companion loses $\sim90\%$ of its mass and evolves off the MS. The BH accretes about $1.5\msun$.

As far as the \rlmxbb\ route is concerned (Fig.~\ref{fig:rlmxb}, lower plot), the initial masses on ZAMS are much smaller and in a typical case equal $\sim23\msun$ and $\sim3.3\msun$. The initial separation is large ($a\approx9400\rsun$). Similarly to \rlmxba\ the primary evolves off the MS and commences a CE phase while being a $\sim10\msun$ CHeB star. The outcome is a compact binary ($a\approx40\rsun$) composed of a $\sim7.9\msun$ HeS and $\sim3.5\msun$ MS star. The primary is not heavy enough to form a BH. Instead, in $100\kyr$ forms a heavy ($\sim1.8\msun$) NS with a strong NK. The orbit becomes highly elongated ($a\approx200\rsun$, $e\approx0.96$). The secondary, which is now the more massive star, expands due to nuclear evolution and after $\sim200\myr$ fills its Roche lobe during periastron passage and a MT begins. In such a situation, we assume that the orbit is immediately circularized. In $\sim500\myr$, the NS accretes $>0.7\msun$, the critical mass is reached, and it collapses to a low-mass ($2.5\msun$) BH. Afterwards, the MT restarts when the secondary is evolving off the MS. Finally, the BH grows to a mass of $\sim3\msun$.

\begin{figure}[t]
    \centering
    \includegraphics[width=1.0\columnwidth]{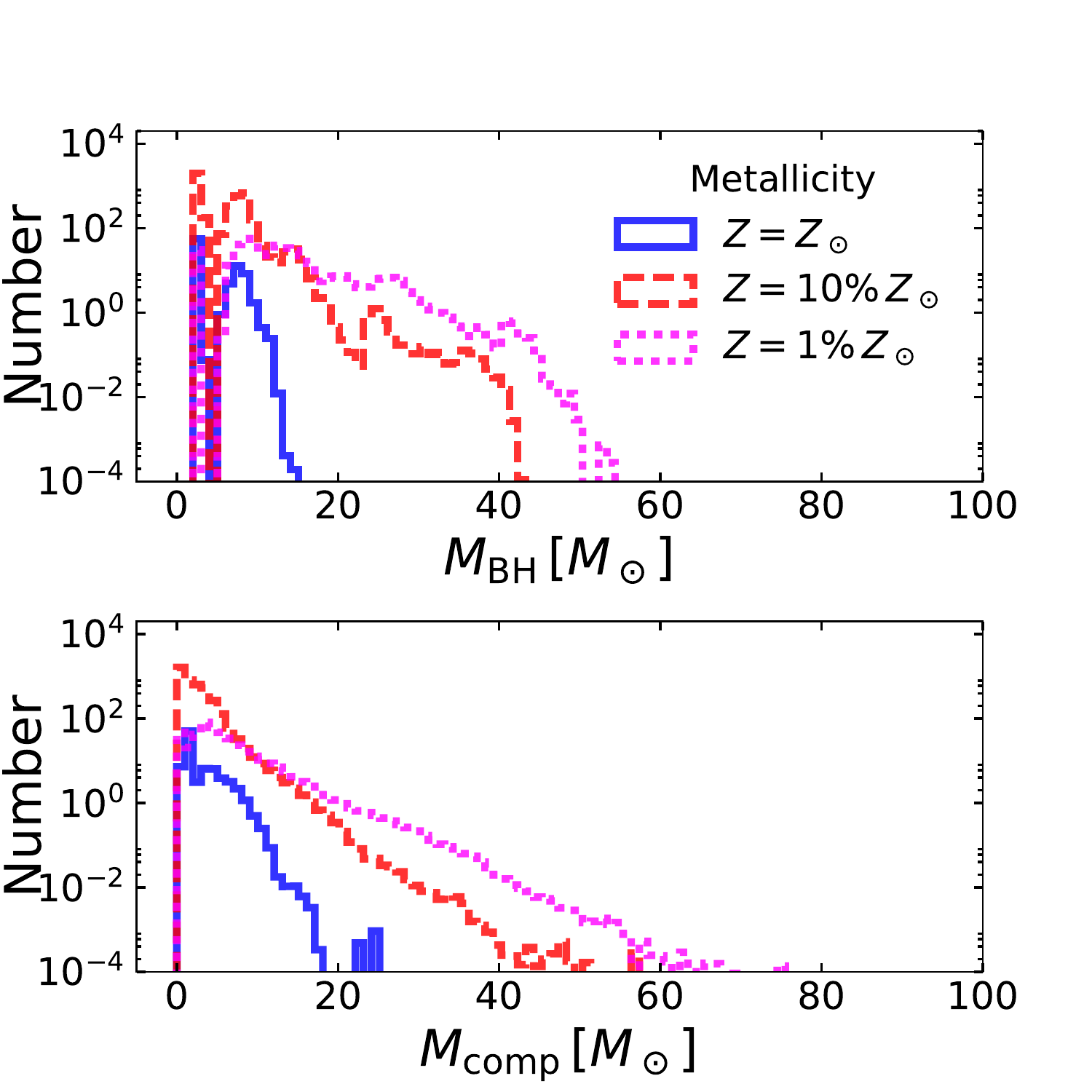}
    \caption{Distributions of BH masses (upper plot) and companion masses (lower plot) in mass transferring (RLOF) binaries for three tested metallicities: $Z=\zsun$ (\std\ model), $Z=10\%\,\zsun$ (\midz\ model), and $Z=1\%\,\zsun$ (\lowz\ model).}
    \label{fig:mt_mass}
\end{figure}

Fig.~\ref{fig:mt_mass} shows distribution of masses for BHs and companions. The leftmost peak (both for BHs and companions distributions) is composed of BHs formed through \rlmxbb\ route, whereas, heavier BHs and companions represent \rlmxba\ route. The maximal BH masses, naturally, exceed these for wide BHBs, or expected from single star evolution \citep{Belczynski1005}. The heaviest companions are usually accompanied by the heaviest BHs what allows to avoid extreme mass ratios, which in our calculations are assumed to lead to dynamical instability during MT.

Models \midz\ and \lowz\ predict larger populations of MTBHBs than \std\ model. In lower metallicity environments it is easier to produce heavier BHs, thus companions Roche lobes are relatively smaller. Also, the nuclear expansion is slower what allows for longer phases of stable MT.

Typical initial center-of-mass velocities ($v_{\rm COM,0}$) are smaller than $20$ km/s. It is a result of the fact that both heavy BHs ($M\gtrsim40\msun$) and heavy NSs ($M>1.8\msun$) present in MTBHBs formed through channels \rlmxba\ and \rlmxbb, respectively, and have low, or negligible NKs. The exception are models with higher average NKs (\nkr\ and \nkbe) which give much higher $v_{\rm COM,0}$ ($50$--$100$ km/s). We note, that velocities around $100$ km/s are also attainable in models with standard NK distribution (e.g. \std) in systems with lighter BHs ($M_{\rm BH}\approx6\msun$).

\subsection{Mergers}\label{sec:mergers}

\begin{figure}[t]
    \centering
    \includegraphics[width=1.0\columnwidth]{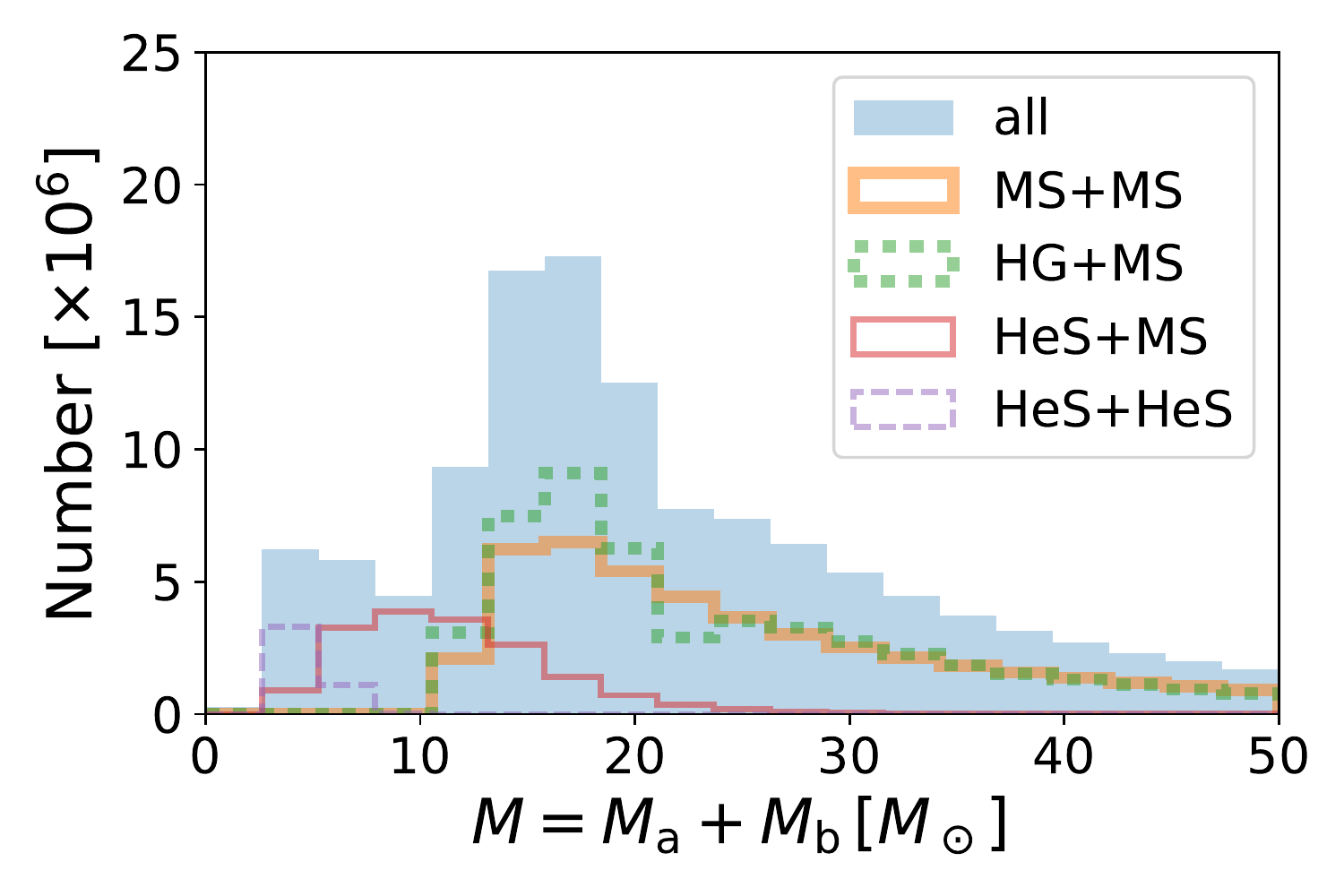}
    \caption{Distribution of the total binary masses prior to merger for \std\ model. Only main types of involved binaries are presented separately. The high-mass tail extends monotonically to $\sim215\msun$. Designations stand for: MS - main sequence; HG - Hertzsprung gap; HeS - helium star.}
    \label{fig:mer_mass}
\end{figure}

Mergers are a frequent outcome in population synthesis calculations of isolated binaries due to a failed CE ejection. Our results show that most typically (more than $95\%$ of cases) mergers occur when the components are on MS, HG, or are HeS (Fig.~\ref{fig:mer_mass}). Although our results are not representative for the entire binary star population, as we include only binaries with initial primary's mass in the range $10\dash150\msun$, we include all potential progenitors of SBHs from merging binaries for which $\mzamsa+\mzamsb\geq20\msun$. We note that even if some lower-mass binaries may produce a BH predecessor after a merger, the poorly understood merger physics does not allow to include such cases in the following analysis. Throughout the paper we use a designation mSBH to distinguish SBHs formed from merger products from these originating from disrupted binaries.

In order to include merger products in the BH population we have implemented a simplified formalism of Olejak et al. (in prep.; see Sec.~\ref{sec:methods_merger}). After merger, the products were evolved as single stars till the formation of a compact object, or reaching the age of $15\gyr$. This way, we were able to roughly estimate the population of mSBHs. According to our simulations, the minimal ZAMS mass that produce a BH in single-star evolution is between $19$--$22\msun$ depending on the metallicity. Similarly, the minimal zero-age helium main-sequence mass that produce a BH is $\sim9\msun$ \citep[e.g.][]{Woosley1901}. In a typical case (MS+MS and HG+MS mergers) there are no interactions prior to merger.

The predicted numbers of BHs for all tested models are provided in Tab.~\ref{tab:results}. The model dependence is very low with \sci{\sim3.6\dash6.6}{7} BHs originating from mergers, except \imff\ and \imfs\ models where the contrast is significant (\sci{1.1}{8} and \sci{1.6}{7}, respectively) due to different (higher, or lower, respectively) number of massive stars on ZAMS.

\begin{figure}[t]
    \centering
    \includegraphics[width=1.0\columnwidth]{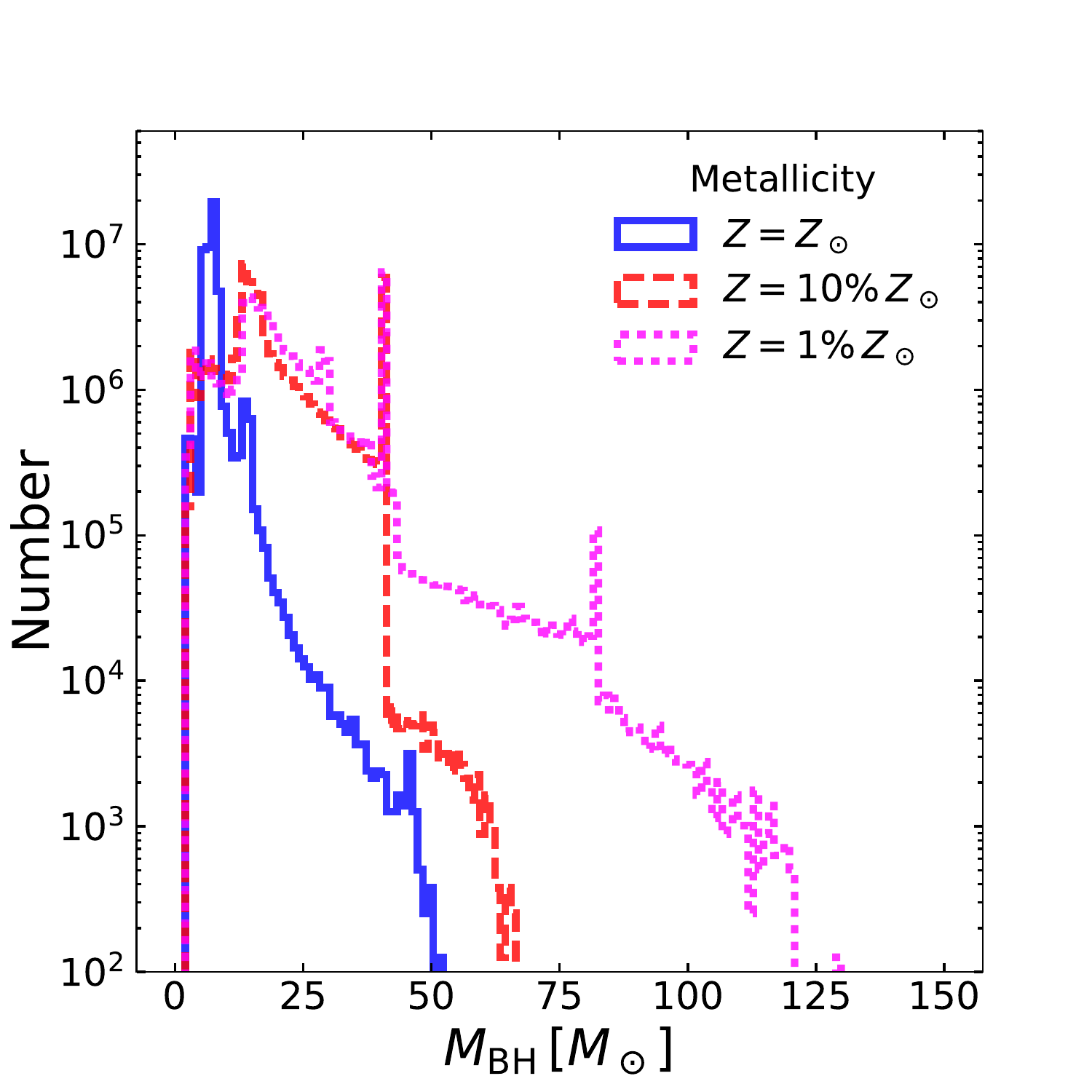}
    \caption{Distribution of BH masses from post-merger products according to the model by Olejak et al. (in prep.; see Sec.~\ref{sec:methods_merger}). A notable feature is the presence of BHs inside the Mass Gap ($2.5$ -- $5\msun$) and BHs much heavier than typical masses obtained through other evolutionary channels (e.g. $\mbh>40\msun$ for \std\ model, or $\mbh>70\msun/120\msun$ for \midz/\lowz\ models. }
    \label{fig:mer_bh_mass}
\end{figure}

The mass distribution (Fig.~\ref{fig:mer_bh_mass}) shows a major peak at $\sim7.5\msun$ ($\sim15\msun$ for \midz\ and \lowz\ models). The distribution noticeably differs from distributions of BH masses obtained through other formation channels. Particularly interesting is the presence of BHs inside the Mass Gap ($\sim2.5$ -- $5\msun$) were no compact objects have been detected through observations \citep[e.g.][]{Bailyn9805,Ozel1012,Farr1111}. These BHs mainly originate from mergers of NSs with HeS thus exceeding the maximal mass for a NS ($M_\mathrm{max,NS}$). In a typical case, the components masses on ZAMS are $\sim10\msun$ and $\sim9.6\msun$. The primary, being slightly heavier, evolves faster and in $\sim24\myr$ fills its Roche lobe while expanding on a HG. The mass transfer leads to a mass reversal with primary becoming a $2.2\msun$ HeS, whereas secondary grows to $14\msun$ still remaining on a MS. In about $4\myr$ primary explodes as an electron capture supernova and forms a NS with a mass of $1.26\msun$ and negligible NK. The separation is $770\rsun$ at the moment. Secondary needs additional $1\myr$ to fill its Roche lobe during a CHeB phase. The mass ratio is extreme with the donor being ten times heavier then the accretor, what leads to a CE and, in a consequence, merger of both stars. Finally, according to our prescription a $2.9\msun$ BH forms. Such low-mass free-floating BHs can be detectable by microlensing methods \citet[e.g.][]{Wyrzykowski2015}. We note that the formation of low-mass BHs through this channel heavily depends on our assumptions concerning mergers (see Sec.~\ref{sec:methods_merger}) and, particularly, the fraction of the HeS which is actually accreted onto the NS.

Another striking feature of the distribution is the presence of high mass tails exceeded up to $\sim50\msun$, $\sim80\msun$, or $\sim130\msun$ for \std, \midz, and \lowz\ models, respectively (Fig.~\ref{fig:mer_bh_mass}). Typically, in \std\ model, progenitors of such massive BHs are formed from a merger of a $\sim25\msun$ HeS with a $46\msun$ MS star which leaves a $\sim43\msun$ HeS as an outcome. Such a massive HeS collapses directly to a BH with $\sim40\msun$ mass. Evolution for massive BHs in other metallicities are analogous. These massive stars, although rare, may produce significantly stronger microlensing events than those with typical masses ($\sim7$--$8\msun$). W note that \citet{Spera1905} obtained even heavier single BHs as a result of stellar mergers using different prescription for post-merger stellar parameters. In contrast to our results, their heavies BHs (up to $\sim145\msun$ for metallicity $Z=0.0001$) originate from mergers of two MS stars.

BHs originating from mergers will have typically very low initial velocities (\ivbh). According to our results more than $75\%$ of the post-merger BHs will have no additional velocity to that resulting from the movement in the galactic potential ($\ivbh\approx0$). Only BHs originating from double NS mergers may obtain high \ivbh\ (above $100\kms$), however, they constitute only $<0.1\%$ of all post-merger BHs. We note, that asymmetries in the merger process and mass ejections may be a source of significant NKs, but are not involved in this study.

\section{Application: Microlensing of bulge stars by isolated BHs}\label{sec:ml}

Gravitational microlensing effect is an amplification of the light of a distant object by another object (luminous, or not) aligned in the line-of-sight toward the source \citep{Paczynski1996}. Unfortunately, lens mass measurement, which is necessary to conclude on its stellar, or BH nature, is typically prone to parameter degeneracies \citep[e.g.][]{Wozniak2001, Sumi2013, Wyrzykowski2015, Mroz2017} and BH lenses can be found only in particular circumstances \citep[e.g.][]{Bennett2002, DongBH, Wyrzykowski1605, Rybicki1805}. On the other hand, population studies allow us to estimate the expected number of microlensing events with BH lenses, what we do in the following analysis. Other similar attempts include studies by \cite{Oslowski2008} and \cite{Lu2019}.

Due to the high star number density and relative proximity, the Galactic bulge is a frequent direction of observations aimed at detecting microlensing events \citep[e.g.][]{Gould0006, Wyrzykowski2015}, therefore, here we also concentrate on this particular direction.

We make a series of simplifying assumptions. First of all, we assume that the source is always located in the bulge, which is represented by a flat disk of size equal $\theta_{\rm bulge}=31$ square degrees, possessing $N_{\star,\rm bulge}=1.5\times10^8$ stars. The tangential velocity distribution (radial velocity can be ignored in microlensing) is on average zero and has a dispersion $\sigma_{\rm bulge,z}=\sigma_{\rm bulge,y}=80$ km/s in both directions: parallel (y-axis) and perpendicular (z-axis) to the galactic plane \citep{Skowron2011}.

Secondly, we assume that a BH lens is always located in the galactic disk, which has a total mass of $5\times10^{10}\msun$ \citep{Licquia1506} and its stellar number density (we assume that BHs have the same distribution as stars in general) is described by 

\begin{equation}\label{eq:rho_star}
    \rho_{\rm star}\propto e^{\frac{|z|}{0.3{\rm kpc}}-\frac{R}{2.6{\rm kpc}}},
\end{equation}
\noindent where $z$ is the distance from the galactic plane and $R$ is the distance from the galactic centre in the galactic plane, following \citet{Batista2011} and \citet{Skowron2011}. 
We assume that components of the tangential velocity of stars in the disk are on average $v_{\rm disk,mean,z}=0$ in the direction perpendicular to the galactic plane and $v_{\rm disk,mean,y}=200$ km/s in the galactic plane. The tangential velocity dispersion is assumed to be $\sigma_{\rm disk,z}=40$ km/s perpendicular to the galactic plane and $\sigma_{\rm disk,y}=55$ km/s parallel to galactic plane.

In contrary to other studies, here we use the BHs from disrupted binaries and stellar mergers only, as BHs in binaries represent small fraction ($<20\%$) of the total population (Tab.~\ref{tab:results}). Additionally, microlensing by binary lens is more complex (caustic crossing, high amplification, etc.) and will be investigated in a separate study. Also, due to the assumed $100\%$ binary fraction for stars heavier than $10\msun$ on ZAMS, we do not expect any BHs formed through single-star evolution.

Stellar mergers may potentially constitute a significant part of the BH population (see Sec.~\ref{sec:mergers}), so it is necessary to include them in the calculation of the microlensing rate. In such a case, not only the number of BHs is important, but also their masses and proper motions. We calculate the final BH mass following the formulas of Olejak et al. (in prep.; see Sec.~\ref{sec:methods_merger} for the post merger mass and evolutionary phase). As far as the proper motions are concerned, we assume that the \ivbh\ of the post merger star is equal to the pre-merger center-of-mass velocity of the binary, i.e. merger occurs without a kick.

We use the Monte Carlo method to sample the spacial and velocity distributions of both lenses and sources. In our analysis we include only lenses which were localized between the observer, assumed to be located at $x_\odot=8.3$ kpc from the Galactic centre, and the Galactic bulge, ignoring less likely lenses and sources from the far Galactic disk.

%einstein radius, angular!
An angular Einstein radius of a lens depends on its mass ($M_\mathrm{L}$) and distances between the observer and the lens ($D_\mathrm{L}$), or the source ($D_\mathrm{S}$):
\begin{equation}
    \theta_\mathrm{E} = \sqrt{\kappa M_\mathrm{L} \pi_\mathrm{rel}}, ~~~~ \pi_\mathrm{rel} = \frac{1}{D_\mathrm{L}} - \frac{1}{D_\mathrm{S}},
    \label{eq:thetae}
\end{equation}
where $\kappa=\frac{4G}{c^2}\approx 8.1 ~ \frac{\mathrm{mas}}{\msun}$. 
If the relative proper motion between the lens and the source is $\vec{\mu}_\mathrm{rel} = \vec{\mu}_\mathrm{L}-\vec{\mu}_\mathrm{S}$, the source crosses the Einstein radius in time called Einstein Radius crossing time, or simply, event's time-scale, $t_\mathrm{E} = \frac{\theta_\mathrm{E}}{|\mu_\mathrm{rel}|}$.

%relative velocity
The relative proper motion is computed using the velocities of the lens and the source as:
\[
    \mu_{\rm rel,y/z}=\frac{v_{\rm rel,y/z}}{x_\odot-x},
\]
and
\begin{multline*}  
    v_{\rm rel,y/z}=v_{\rm BH,y/z}-v_{\rm Earth,y/z} \\
    -(v_{\rm source,y/z}-v_{\rm Earth,y/z})(1-\frac{x}{x_\odot}),
    %(v_{\rm BH,y/z}-v_{\rm Earth,y/z})/D_{\rm{L}}-(v_{\rm source,y/z}-v_{\rm Earth,y/z})/D_{\rm{S}}),
\end{multline*}

\noindent where $v_{\rm rel,y/z}$ is a component of the relative velocity along y-/z-axis. $v_{\rm BH,y/z}$, $v_{\rm source,y/z}$, and $v_{\rm Earth,y/z}$, are the velocities of the BH, the source, and the observer, along y-/z-axis, respectively. The observers velocity was assumed to be $v_{\rm Earth,y/z}=230/15.5$ km/s. $x=x_\odot-D_{\rm L}$ is the distance of the lens from the Galactic centre. 

Finally, we calculated an estimated number of microlensing events (${\rm E}(N_{\rm ML})$) during $t=1$ yr as (assuming $\theta_{\rm E}$ is small), 
\[
    {\rm E}(N_{\rm ML})=\frac{\sum_i (2\theta_{\rm E,i} \frac{v_{\rm rel,i}}{x_\sun-x_{\rm i}} t + \pi \theta_{\rm E,i}^2)}{\Omega_{\rm bulge}}N_{\star, bulge},
\]
\noindent where, $\theta_{\rm E,i}$ is the Einstein radius generated by i-th BH, $v_{\rm rel,i}=\sqrt{v_{\rm rel,x,i}^2+v_{\rm rel,y,i}^2}$ is the relative tangential velocity of the i-th BH, and $\Omega_{\rm bulge}$ is the considered area of the bulge in steradians. The summation goes over all the included BHs, i.e localized between the observer and the bulge.

If we assume that the stellar mass of the Galactic disk is $\sim5\times10^{10}\msun$ \citep{Licquia1506} and we take the number of stars in the bulge as $N_{\star, bulge}= 1.5\times 10^8$ from the number of monitored stars by the OGLE-III survey covering 31 sq.deg of the bulge \citep{Wyrzykowski2015}, we predict that there might be as many as $\sim14$ yr$^{-1}$ microlensing events due to BHs.
In the so-far analysed OGLE-IV data \citep{Udalski2015, Mroz2019}, which covers $160$ sq.deg and monitors $400$ million sources, the number of expected events due to BHs grows to $\sim26$ yr$^{-1}$. 
Here we include only events with characteristic time $t_\mathrm{E}$ in range $1$--$1000$ days which includes most of the events ($>95\%$). Although BHs from disrupted binaries have higher velocities and, therefore, higher chance for microlensing, BHs from disrupted binaries (dSBH; Sec.~\ref{sec:SBH}) dominate the rate ($\sim55\%$ of events) due to higher masses and greater abundance.

\begin{figure}
    \centering
    \includegraphics[width=1.0\columnwidth]{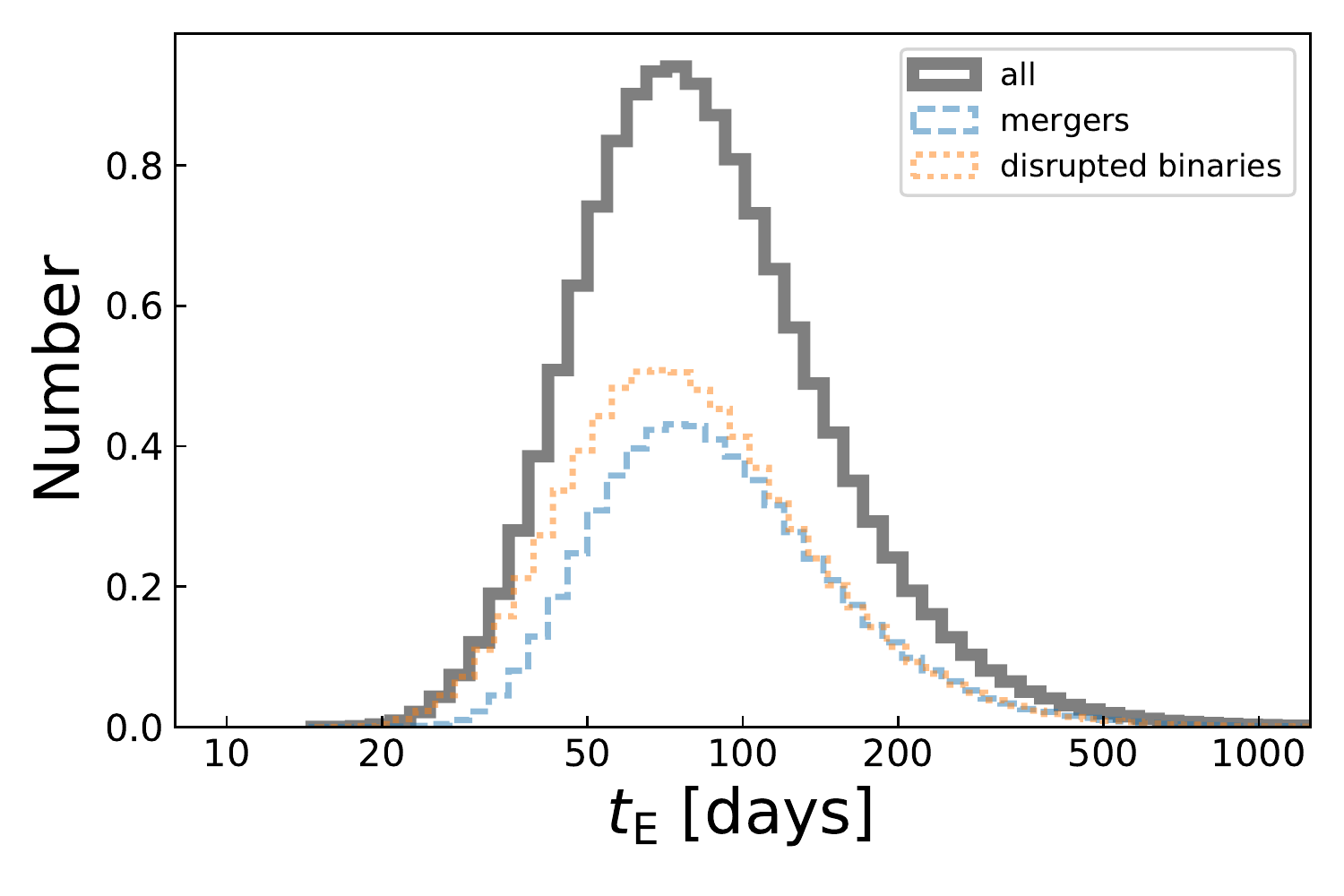}
    \caption{Distribution of microlensing time-scales (Einstein ring crossing times, $t_\mathrm{E}=r_\mathrm{E}/v_\mathrm{rel}$) for BH lenses from binaries. We make a division on BHs originating from stellar mergers and these from disrupted binaries. The average values of $t_\mathrm{E}$ for BHs from merger ($\sim108$ days) are relatively higher than for BHs from disrupted binaries ($\sim97$ days).} 
    \label{fig:t_E}
\end{figure}

For a given source and lens distances and their velocities we computed $\theta_\mathrm{E}$ and $\mu_\mathrm{rel}$ and hence the distribution of time-scales ($t_\mathrm{E}$, Fig.~\ref{fig:t_E}). The typical values of $t_\mathrm{E}$ are on average $\sim97$ days for BH originating from disrupted binaries (dSBH; Sec.~\ref{sec:SBH}) and $\sim108$ days for these originating from stellar mergers (mSBH; Sec.~\ref{sec:mergers}. 

Events with such time-scales are detectable in current microlensing surveys with efficiency of more than $75\%$ \citep[e.g.][]{Sumi2013, Wyrzykowski2015}. This means in the OGLE-III data covering 8 years of monitoring of $150$ million of stars there should be about $84$ events due to BHs. Since, as shown above, these tend to have longer time-scales, the annual Earth motion (the parallax effect) is likely to cause the distortion to the standard microlensing light curve, therefore, it is possible that some fraction of those $84$ events are not in the \citet{Wyrzykowski2015} sample of standard events. Indeed, \citet{Wyrzykowski1605} have identified $59$ OGLE-III events with significant parallax effect, among which $13$ are likely due to NSs and BHs. The improved analysis of \citet{WyrzykowskiMandel2019} has increased the number of NS and BH lenses to $18$ in that sample. The remaining undetected BH lenses might still await discovery in the OGLE-III data with moderate and small annual parallax signatures.

\section{Discussion}\label{sec:discussion}

\subsection{Detectability of single BHs}

Although the predicted number of SBHs in MWEG (both dSBH and mSBH) is very significant (\sci{3.2}{7} -- \sci{2.5}{8}), these objects still evade detection. Already \citet{Shapiro83} estimated that BHs in the Galaxy should be counted in millions. \citet{Timmes9602} estimated that there are even $1.4\times10^9$ BHs in the Galaxy. \citet{Samland9803} calculated the BH population to be $1.8\times10^8$. They included the changes in SFR, but excluded binary stars from their analysis. It was proposed that SBHs may be detected when interacting with interstellar matter \citep{Shvartsman7112,Agol0208,Barkov1211,Tsuna1801,Tsuna1907}, however, the estimated X-ray emission from an accreting SBH is very small \citep[$<10^{35}\ergs$; e.g.][]{Barkov1211}. Such faint sources may be potentially detected only in our Galaxy. \citet{Matsumoto1803} suggested that the accretion onto a SBH from interstellar medium may not be necessarily spherical and result in the formation of an accretion disk. In such disks the same instabilities may arise as in X-ray Novae resulting in a transient behavior with outburst luminosities reaching $\sim10^{38}\ergs$.

SBHs can be also detected as a gravitational microlenses \citep{Agol0209,Wyrzykowski1605,Rybicki1805}. As shown in this paper, SBHs may have very high velocities what increases a chance for microlensing event. On the other hand, a significant part of DCOs (\rdcob\ channel) have separations small enough to act as a single, very massive lens, with the mass as the sum of the masses of individual components. In combination with small proper motions of such binaries, they may generate long-lasting (t$_{\rm{E}}\sim1\yr$) events, however, we note that undisrupted compact DCOs make less than 15\% of all BHs in the Milky Way.

Although none SBHs were detected so far, a few observational methods were proposed. The predicted number of observations in future X-ray missions is strongly dependant on the BH velocities and accretion models. Future missions may help to improve the constrains on this physics. Nevertheless, detectability of SBHs in X-rays is strongly spoiled by background AGNs, hard coronal emitters and CVs \citep{Motch12}. Another possibility, to which more attention is put in this work, is the detection of SBHs with microlensing \citep{Paczynski8605}. First candidates were proposed in \cite{Bennett2002} and \cite{Mao2002} and were followed-up with X-ray observations, however, no signal from ISM accretion was detected \citep{Maeda2005, Nucita2006}. Recently, several candidates were proposed \citep{Wyrzykowski1605}, but the lens's mass estimates are heavily degenerated due to the lack of measurement of relative velocities. Consequently, WDs or NSs cannot be fully ruled out.  For the currently on-going events, there is an opportunity for degeneracy breaking thanks to astrometric measurements from optical interferometry \citep{Dong2019} or Gaia\citep{Rybicki1805}, however, this will only be possible for the brightest events. In near future with the LSST it should be possible to observe thousands of microlensing events due to BHs and, therefore, to confront the stellar evolution predictions with the observed parameters of Galactic BHs.

\subsection{Detectability of BHs in binaries}

All dynamically confirmed galactic BHs were detected in XRBs. We note that a BH candidate was also detected  in a non-interacting binary \citep{Thompson1806} and many more are expected to be seen by Gaia \citep{Mashian1709,Breivik1711,Yamaguchi1807}. In XRBs a compact object accretes mass from its companion which results in a formation of an accretion disk and production of highly energetic radiation. If the companion is observable, radial velocities can be measured and the mass function can be estimated. Unfortunately, the majority of XRBs containing BHs are transient systems, thus are visible only during outbursts and only a few recurring systems were observed. What is more, typical transient systems are characterized by low mass donors ($M_\mathrm{don}\lesssim2\msun$), therefore, undetectable from extra-galactic distances and outshined by the disk during an outburst. In $19$ XRBs, BHs were dynamically confirmed \citep[e.g.][]{Wiktorowicz1409}, however, there are many more candidates \citep{Corral-Santana1603}.  

Gravitational waves emission from double BH mergers may be perceived as a detection of BHs. Up-to-now, 10 binary BH merger events were detected by LIGO/Virgo \citep{Abbott1811}. Using different pipeline \citet{Venumadhav1904} found 6 additional sources in publicly available data from the second observing run. Notably, in spite of one double NS merger detection \citep{Abbott1710}, no BH+NS mergers were detected. The masses of pre-merger BHs range from $\sim7.7\msun$ to $\sim50.6\msun$ with the total binary mass as high as $\sim85.1\msun$ \citep{Abbott1811}. Many tens of detections are expected from the next observational runs \citep{LIGO1811} and several candidates have already been found \citep[e.g.][]{Singer1904}.

BHs in binaries may be detected also through microlensing. They may lens differently depending on the compactness of the binary. If the separation is significantly smaller than the total mass Einstein radius, the system will act as a single lens with a mass equal to the sum of the components' masses. On the other hand, if the system is very wide (many Einstein radii), a BH will act as a single lens unaffected by a remote companion. In-between these two options, we have systems that act as a binary lens with all the effects connected with caustics crossing. Up-to-now, no microlensing events with BH lenses were confirmed. Although there are several candidates \citep[e.g.][]{Miyake2012, Shvartzvald2015, Wyrzykowski1605}, parameter degeneracies make it impossible to infer mass, thus distinguish between a BH and other low-luminosity objects like WDs.

\subsection{Comparison with previous estimates of microlensing rates by BHs}

Our prediction of $\sim14$ microlensing events due to BHs in a year stays in contrast to previous estimates. \citet{Gould0006} analytically calculated that only $1\%$ of microlensing events in the direction of the Bulge are due to BH lenses, what gives about 30 events in OGLE-III (3-4 per year). There are several reasons for this contradiction. First of all, \citet{Gould0006} concentrates on BH lenses in the bulge, whereas our lenses are in the disk. $\pi_{\rm rel}$ (see Eq.~\ref{eq:thetae}) is typically smaller for bulge lenses, thus also the Einstein radius, which is proportional to the probability of microlensing. Secondly, \citet{Gould0006} assumed $40\msun$ as a minimal ZAMS mass producing a BH remnant, whereas in our calculations we take $\sim22\msun$ (for solar metallicity). Consequently, they obtained much less BHs from the same initial population. Moreover, BHs in \citet{Gould0006} follow the velocity distribution of stars, whereas in our simulations BHs obtain additionally a kick (including NK, Blaauw kick, and the kick from binary disruption). Thirdly, the binary systems were ignored in the work of \citet{Gould0006}, while here we show that the binary systems could potentially contribute significantly to the population of lensing BHs.

A more recent work, \citet{Rybicki1805} used the same population synthesis \startrack\ code to estimate the microlensing rate events due to BHs. However, they focus on astrometric microlensing events, i.e. events in which not only photometric observation is possible, but also astrometric signal can be detected with {\it Gaia} space mission. They estimated there should be a few such events per year. About an order of magnitude of difference between our results stems from the fact that only $\sim5\%$ of bulge stars ($5\times10^6$ in their work) are bright enough to potentially generate a detectable astrometric microlensing event. In contrast, we included all bulge stars as potential sources ($1.5\times10^8$), thus $30$ times more, which results in significantly higher estimated rate. Moreover, \citet{Rybicki1805} used averaged, or typical values to calculate the rates, whereas in this paper we populate the Galaxy with stars, evolve them to the formation of a BH and calculate the rates directly.

We note that in our approach to estimate the microlensing rates we made a number of simplification. Especially, we used only one model from 8 analysed in this papers. Additionally, we chose a simple model of BH distribution in the Galaxy. For example, model of \citep{Robin0310}, which was used in \citet{Rybicki1805}, gives microlensig rate by BH of $\sim9$ yr$^{-1}$, so nearly two times smaller than the rate obtained for the distribution described by Eq.~\ref{eq:rho_star}. We also note that the rate may change by a factor of $\sim5$ depending on the chosen evolutionary parameters, e.g. IMF, natal kicks, metallicity. 

\subsection{Hertzspurng Gap donors in common envelope evolution}\label{sec:modelsAB}

\begin{deluxetable*}{lcrlrlrlrl}
    \tablewidth{\textwidth}
    \tablecaption{Number of BHs in the Milky Way Equivalent Galaxy (model A)}

    \tablehead{ Model & $N_{\rm BH,tot}$ & \multicolumn{2}{c}{$N_{\rm dSBH}$} & \multicolumn{2}{c}{$N_{\rm BHB}$} & \multicolumn{2}{c}{$N_{\rm BH,DCO}$} & \multicolumn{2}{c}{$N_{\rm mSBH}$}}
    \startdata
    \std  & \sci{1.0}{8} & \sci{5.4}{7} & ($54.2\%$) & \sci{1.5}{6} & ($1.5\%$) & \sci{6.1}{6} & ($6.1\%$) & \sci{3.8}{7} & ($38.2\%$) \\
    \midz & \sci{1.2}{8} & \sci{3.5}{7} & ($29.9\%$) & \sci{2.0}{6} & ($1.7\%$) & \sci{2.2}{7} & ($18.8\%$) & \sci{5.8}{7} & ($49.6\%$) \\
    \lowz & \sci{1.2}{8} & \sci{3.7}{7} & ($30.7\%$) & \sci{1.5}{6} & ($1.2\%$) & \sci{2.1}{7} & ($17.4\%$) & \sci{6.1}{7} & ($50.6\%$) \\
    \ssa  & \sci{1.0}{8} & \sci{6.6}{7} & ($65.0\%$) & \sci{3.4}{6} & ($3.3\%$) & \sci{8.1}{6} & ($8.0\%$) & \sci{2.4}{7} & ($23.6\%$) \\
    \nkr  & \sci{1.0}{8} & \sci{6.4}{7} & ($63.8\%$) & \sci{6.0}{4} & ($0.1\%$) & \sci{2.1}{5} & ($0.2\%$) & \sci{3.6}{7} & ($35.9\%$) \\
    \nkbe & \sci{1.0}{8} & \sci{6.5}{7} & ($64.3\%$) & \sci{5.5}{4} & ($0.1\%$) & \sci{8.7}{4} & ($0.1\%$) & \sci{3.6}{7} & ($35.6\%$) \\
    \imff & \sci{2.5}{8} & \sci{1.4}{8} & ($56.0\%$) & \sci{3.2}{6} & ($1.3\%$) & \sci{1.6}{7} & ($6.4\%$) & \sci{9.1}{7} & ($36.4\%$) \\
    \imfs & \sci{3.1}{7} & \sci{1.6}{7} & ($50.9\%$) & \sci{5.6}{5} & ($1.8\%$) & \sci{1.9}{6} & ($6.0\%$) & \sci{1.3}{7} & ($41.3\%$) \\
    \enddata
    \tablecomments{The same as Tab.~\ref{tab:results}, but for a models in which HG donors were allowed to survive the CE phase (see Sec.~\ref{sec:modelsAB} for details).} 
\label{tab:results_modelA}
\end{deluxetable*}

It is still not known if binaries undergoing CE events with HG donors will merger due to the lack of clear core-envelope boundary, or they can survive, because the boundary develops during the HG phase. In order to check the importance of our assumption (model B - CE with HG donor always leads to a merger), we have tested also an opposite case (model A) where binaries are allowed to survive the CE phase even if donor is on its HG. The results are presented in Tab.~\ref{tab:results_modelA}.

The difference in the total number of BHs is small ($\lesssim10\%$), as most of the BHs form without a CE phase during the binary evolution. More pronounced is the higher number of DCOs (particularly, the close ones), which may exceed two orders of magnitude in comparison to model B for higher NK models (\nkr\ and \nkbe). A part of close DCOs progenitors evolve through a CE phase and in model A donors are frequently HG stars.

The number of mergers leading to the formation of a BH is also affected. In model B, CE events with HG donors are treated as mergers, what is not always a case in model A. In a consequence, there are significantly fewer (up to $33\%$) BHs originating from stellar mergers in model A than in model B.

As the total number of BHs is affected only slightly by the treatment of HG donors in CE events, we predict a negligible impact on the microlensing observations. However, it may be significant if events caused by close systems are considered.

\subsection{Binary fraction}\label{sec:binary_fraction}

In this work, we assumed that all stars heavier than $10\msun$ are born in binaries. Such a high binary fraction ($100\%$) for massive stars is supported by observations, which suggest their binary fraction to be higher than $90\%$ \citep{Sana1207}. For lower mass stars we assumed equal number of binaries and single stars on ZAMS (binary fraction equal $50\%$).

Nonetheless, we agree that the binary fraction may not change in such a drastic way when the mass of the primary increases. More probably, the binary fraction increases continuously. Therefore, a fraction of massive stars, which are heavy enough to produce BHs ($M_{\rm ZAMS}\gtrsim20\msun$) may actually form without companions. If included in our simulations, these stars could increase the relative fraction of single BHs in the entire BH population. However, the total number of BHs is expected to be lower due to the fact that flat mass ration distribution in binaries gives higher average mass of two stars in a binary, thus more BH progenitors, than two single stars following the IMF relation. We note, that mergers of binary systems of two massive stars (both heavy enough to form a BH in single stars evolution), may decrease the total number of BHs, but mergers of NS progenitors ($M_{\rm zams}\approx8$--$20\msun$) may become heavy enough to form a BH in posterior evolution, thus increasing the total number of BHs.

Many of the massive binaries may be part of triples and higher-order systems \citep[e.g.][]{Toonen1612}. Even $\sim50\%$ of massive (OB type) stars may exist in triples \citep[e.g.][]{Sana1411}, but the inner binary in a hierarchical triple system may evolve effectively isolated from the third star, thus estimating the significance of higher-order systems on BH populations is complicated. Of particular interest are mergers induced by Kozai-Lidov mechanism in stellar triples \citep[e.g.][]{Antonini1706}. However, higher-order systems are not understood yet well enough to be included in population synthesis modeling.

\section{Summary}
   
In this paper we analysed the general properties of synthetic BH populations in different stellar environments represented by models which differ in the metallicity, initial parameter distributions, and natal kick prescriptions. We note that the results are applicable for further studies like predictions for present and future survey missions, or in-depth analysis of specific BH populations (e.g. the Milky Way galaxy). We particularly focused on BHs originating from disrupted binaries and stellar mergers, which were frequently omitted in previous studies. We find that those BHs constitute a majority of the total BH population.

Particularly, we show that single BHs dominate the total population of BH ($\gtrsim83\%$ of all BHs), even though massive stars form predominantly in binaries. Both binary disruptions and stellar mergers are important and the predicted number of BHs is only slightly affected (up to a factor of $\sim3$) by a chosen model. Although BHs in binaries constitute only a small part ($\lesssim17\%$) of the population, their number is heavily dependent (about two orders of magnitude) on the adopted model parameters (especially natal kick prescription and star formation history), therefore, new observations may significantly help to constrain these parameters and better understand evolutionary processes \citep[e.g.][]{Maccarone1904}.

Using our results, we calculated the expected rate of microlensing events with BH lenses in the direction of the Galactic bulge. We expect that as many as $\sim14$ such events per year with average crossing times around $100$ days in the OGLE-III footprint and about $26$ in the OGLE-IV data. Only some of these events may be observable from Earth due to technical limitation (e.g., low luminosity, extinction). In this estimate we have neglected BHs which remained in binaries, because, low fraction ($\lesssim17\%$ of all BHs) and slow velocities (typically, $\lesssim20$ km/s) allow to suppose that their influence is small.

A grid of 54 models (including 8 main models analysed in the paper) is available in a free-access database\footnote{https://universeathome.pl/universe/bhdb.php}. In our future work we plan to utilize these datafiles in order to study microlensing by single BHs in the Galaxy and provide detailed predictions for surveys like Gaia, or Einstein Telescope.

\acknowledgements
We are thankful to the anonymous referee who helped to improve the paper and thousands of volunteers, who took part in the {\it Universe@Home} project\footnote{https://universeathome.pl/} and provided their computers for this research. We acknowledge the Polish NCN grant HARMONIA 2015/18/M/ST9/00544 to {\L}W, OPTICON H2020 EC grant no. 730890, MNiSW grant DIR/WK/2018/12 and SONATA BIS 2 DEC-2012/07/E/ST9/01360. G.W. is partly supported by the President’s International Fellowship Initiative (PIFI) of the Chinese Academy of Sciences under grant no.2018PM0017 and by the Strategic Priority Research Program of the Chinese Academy of Science Multi-waveband Gravitational Wave Universe (Grant No. XDB23040000). KB and GW acknowledge NCN grant MAESTRO 2018/30/A/ST9/00050. MC and JK acknowledge support from the Netherlands Organisation for Scientific Research (NWO). We acknowledge the Polish NCN grant PRELUDIUM 2017/25/N/ST9/01253 to KR. This work is partly supported by the National Key Program for Science and Technology Research and Development (Grant No. 2016YFA0400704), and the National Natural Science Foundation of China (Grant No. 11690024 and 11873056).

\bibliographystyle{apj}
\bibliography{ms}

\end{document}